\documentclass[aps,notitlepage,twocolumn,nofootinbib,floatfix,pra,10pt]{revtex4-2}
\usepackage{amsmath,amsfonts,amssymb,amsthm,bbm,graphicx,enumerate,times,mathtools,hyperref,physics,marvosym,wasysym}
\hypersetup{colorlinks,linkcolor=blue,citecolor=red,urlcolor=blue}

\usepackage{xcolor}

\definecolor{lightgreen}{RGB}{144, 238, 144} 
\usepackage{csquotes}
\usepackage{float}
\usepackage[ruled,longend]{algorithm2e}
\usepackage{fancyref}
\usepackage{subcaption} 
\usepackage{afterpage} 
\usepackage{placeins} 
\usepackage{caption} 
\usepackage{etoolbox}
\usepackage{mdframed}
\usepackage{subcaption}
\usepackage{tabularx}
\usepackage{appendix,soul}
\usepackage[noabbrev,capitalise]{cleveref}

\begin{document}

\title{beSnake: A routing algorithm for scalable spin-qubit architectures}

\author{Nikiforos Paraskevopoulos$^{1,2}$}
\author{Carmen G. Almudever$^{3}$}
\author{Sebastian Feld$^{1,2}$}

\affiliation{$^{1}$Quantum and Computer Engineering Department, Delft University of Technology, 2628 CD Delft, The Netherlands}
\affiliation{$^{2}$QuTech, Delft University of Technology,  2628 CJ Delft, The Netherlands}
\affiliation{$^{3}$Computer Engineering Department, Universitat Politècnica de València, Camino de Vera, s/n, 46022 València, Spain}


\begin{abstract}

As quantum computing devices increase in size with respect to the number of qubits, two-qubit interactions become more challenging, necessitating innovative and scalable qubit routing solutions. In this work, we introduce beSnake, a novel algorithm specifically designed to address the intricate qubit routing challenges in scalable spin-qubit architectures. Unlike traditional methods in superconducting architectures that solely rely on SWAP operations, beSnake also incorporates the shuttle operation to optimize the execution time and fidelity of quantum circuits and achieves fast computation times of the routing task itself. Employing a simple breadth-first search approach, beSnake effectively manages the restrictions created by diverse topologies and qubit positions acting as obstacles, for up to 72\% qubit density. It also has the option to adjust the level of optimization and to dynamically tackle parallelized routing tasks, all the while maintaining noise awareness. Our simulations demonstrate beSnake's advantage over an existing routing solution on random circuits and real quantum algorithms with up to $1,000$ qubits, showing an average improvement of up to $80\%$ in gate overhead and $54\%$ in depth overhead, and up to $8.33$ times faster routing times.

\end{abstract}

\maketitle


\section{Introduction} \label{Introduction}



The quantum software's ability to optimally transform hardware-agnostic quantum circuits such that they comply with all operational constraints of the architecture -- a process known as mapping -- is at the forefront of quantum compiler development right now \cite{SpinQ,sivarajah2020t,Qiskit,khammassi2021openql,salm2021automating,Developers2023-gs,computing2019pyquil,javadiabhari2014scaffcc,chong2017programming,wu2023intel}. One of the tasks of the mapping process is to account for the circuit's two-qubit interactions and to deal with constraints imposed by the device architectures, notably the limited qubit connectivity. These restrictions are overcome by carefully inserting SWAP gates and thus moving quantum information around the devices -- usually in nearest-neighbor positions -- so that the needed two-qubit gates can be performed. This particular process is also known as qubit routing, and its effectiveness is paramount to maximize the operational fidelity of Noisy Intermediate-Scale Quantum (NISQ) \cite{preskill2018quantum} devices, which are highly prone to different kinds of errors.

SWAP-based qubit routing algorithms have been extensively developed for superconducting quantum devices \cite{tannu2019not,li2019tackling,saki2022error,murali2019noise,niu2020hardware,sinha2022qubit,liu2023tackling,lao2021timing} as these are the most-developed qubit technology when it comes to the number of qubits and system availability in general. Spin-qubit devices, however, are not as mature, and consequently, there are no specialized routing techniques yet that can take advantage of their unique features. The fundamental differentiating aspect of a routing algorithm targeting spin-qubit architectures is the primary method of communication. Here, the \textit{shuttling} operation is used as a means of communication within the device instead of SWAP gates used in superconducting technology. Whereas a SWAP gate exchanges the quantum state of two qubits, a shuttle operation results in the physical relocation of a qubit. Note that SWAP gates can also be implemented in spin-qubit devices; however, shuttling operations are preferred due to their superior operational fidelity and faster execution time. 

Having said that, the shuttle operation brings unique routing challenges that have not been encountered before. A primary example is the need to have at least one, and preferably more, empty sites for the qubits to be able to move, as two qubits cannot be allocated to the same place at the same time. Of course, this is not new for modular ion-trap devices, as they are also using shuttles to move ions between traps. However, due to their fundamental architectural differences with spin-qubit devices, the proposed routing techniques are not applicable \cite{murali2020architecting,saki2022muzzle,schmale2022backend,schoenberger2023using,pozzi2022using}. Additionally, routing for spin-qubit devices deviates from the conventional notion of only enabling two-qubit interactions. It also extends to other types of operations benefiting from the shuttle operation, which necessitates innovative solutions able to handle the different cases. One of these operations is a fast and low-error single-qubit shuttle-based Z rotation, as used in this \cite{li2018crossbar} crossbar architecture proposal.

On that note, a shuttle-based SWAP routing algorithm for a spin-qubit device has been introduced in \cite{SpinQ}. Although it can perform qubit routing in polynomial time in terms of number of qubits, gates, or two-qubit gate percentage, the algorithm is tightly coupled to the strict architectural constraints of a crossbar architecture \cite{li2018crossbar}, making it practically unusable for other architectures. To this date, no routing algorithm has been proposed that fully takes advantage of the shuttle operation, adapts to any architecture, and freely moves qubits around the topology. In this paper, we close this gap by presenting beSnake, a novel routing algorithm for scalable spin-qubit architectures. In addition to the mentioned characteristics, it is also able to simultaneously handle multiple two-qubit gates and shuttle-based Z rotations, thus aiming at a low circuit depth overhead. beSnake's core design prioritizes execution speed, thus making it suitable for high qubit counts. In order to further reduce the computation time, it can be configured accordingly at the expense of slightly higher gate overhead. Other design considerations give beSnake the ability to choose less noisy shuttling paths and, optionally, use SWAP operations, assuming that the architecture supports them.

We have extensively evaluated beSnake by routing random generated quantum circuits that stress-test all its capabilities. Finally, comparisons with the Shuttle-based SWAP (SBS) algorithm \cite{SpinQ} on random and real quantum circuits show on average an up to $80\%$ and $54\%$ improvement in gate overhead and depth overhead, respectively, and up to $8.33$ times faster routing time.

The main contribution of this paper is beSnake, the truly first routing algorithm for scalable spin-qubit architectures. It features the following novel characteristics:
\begin{enumerate}
    \item Utilizing the full freedom of the shuttle operation to move qubits in any direction within a given topology.
    
    \item Efficiently handling complex routing scenarios involving parallelized two-qubit and shuttle-based Z gates. 
    
    \item Adapting to various architecture topologies while being noise-aware and able to adjust the level of optimization.
    
    \item Significantly improving over the state-of-the-art (the Shuttle-based SWAP routing algorithm) with a considerable decrease in gate and depth overhead for up to $1,000$ qubits.
\end{enumerate}

The remainder of this paper is structured as follows: Section \ref{Problem statement} presents the problem statement regarding routing for spin-qubit architectures with particular use of the shuttle operation. In Section \ref{Related work}, we discuss related work and their limitations, including qubit routing techniques used in other quantum technologies as well as algorithms used in classical computing. We present beSnake in Section \ref{The beSnake routing algorithm}, explain its internal functionality with three representative examples, and discuss special routing cases in detail. Then, in Section \ref{Simulations}, we thoroughly analyze the performance of beSnake by considering several configurations related to different levels of shortest path optimizations, the SWAP replacement functionality, the behavior on more connected topologies, and on various qubit densities. In the same section, we also compare beSnake with the Shuttle-based SWAP routing algorithm \cite{SpinQ} on random and real algorithms with up to $1,000$ qubits and discuss the obtained improvements. Finally, we discuss future directions and draw conclusions in Section \ref{Conclusions}.

\section{Problem statement} \label{Problem statement}

\begin{figure*}[htpb]
         \centering
     \begin{subfigure}[b]{0.497\textwidth}
         \centering
         \includegraphics[width=\textwidth]{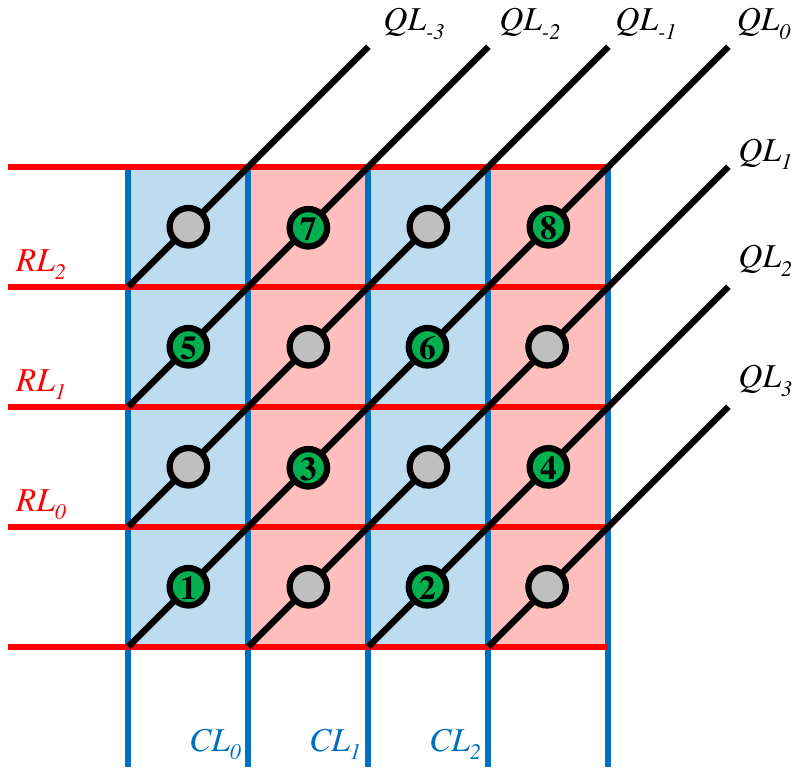}
         \caption{ }
         \label{fig:crossbar_architecture}
     \end{subfigure}
\hfill
     \begin{subfigure}[b]{0.35\textwidth}
         \centering
         \includegraphics[width=\textwidth]{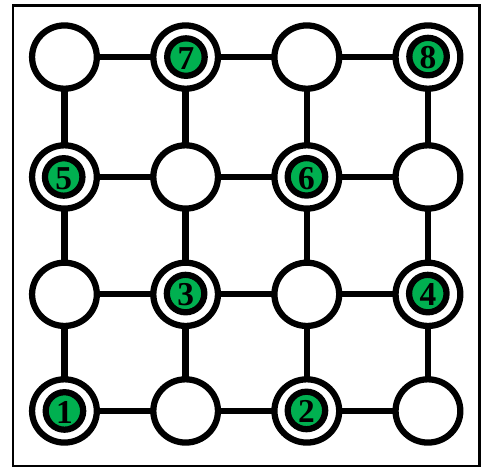}
         \caption{ }
         \label{fig:topologya}
     \end{subfigure}
         \caption{\textbf{(a)} Schematic overview of the crossbar architecture with the various operational control lines (vertical CL, horizontal RL, and diagonal QL). These lines are used with precise pulse sequences and shared among multiple sites, sixteen in this figure, to perform operations on the qubits. Here, the qubits (green circles with numbers) are initialized in a checkerboard pattern. \textbf{(b)} Abstraction of the crossbar architecture representing the coupling between qubits. Each circle represents a quantum dot, and each edge represents a coupling link signifying allowed interactions.}
         \label{topology}
\end{figure*}

Spin-qubit realizations come with unique physical characteristics that make them a promising technology to scale up quantum computing systems \cite{yoneda2018quantum,camenzind2022hole,hendrickx2021four,chatterjee2021semiconductor,RevModPhys.85.961,PhysRevA.57.120,vandersypen2017interfacing,veldhorst2015two,zajac2015reconfigurable,watson2018programmable}. The fundamental building block offering such characteristics is the quantum dot, in which a confined electron or a hole can define a physical qubit \cite{hanson2007spins}. The spin-qubit can then be controlled electromagnetically through several carefully fabricated gate electrodes around it. These electrodes can enable single-qubit or two-qubit operations by precise pulse sequences across various multi-quantum dot arrangements. Such systems have been extensively studied in 1D and, recently, in 2D array formats \cite{hendrickx2021four,borsoi2024shared}. Taking, for instance, the crossbar architecture of \cite{li2018crossbar} shown in Fig. \ref{fig:crossbar_architecture}, we can see the different operational lines and sites where spin qubits can be initialized. In this case, they are initialized in a checkerboard pattern. Two-qubit gates are only performed between two vertically or horizontally adjacent qubits, meaning that this architecture, or any architecture for that matter, only supports nearest-neighbor interactions. Therefore, its topology can be represented by a 2D grid, as shown in Fig. \ref{fig:topologya}. Each node (circle) represents a site of the grid, and each edge connecting the nodes represents the possibility of interaction between the two sites. This type of abstract view is interchangeably called either topology, coupling graph, or layout of the quantum processor.

The problem of bringing the qubits to neighboring positions to execute a two-qubit gate is known as qubit routing. During the NISQ era, it is crucial for the routing process to strive for gate and depth overhead minimization of quantum circuits due to devices suffering from high noise rates. As the number of quantum dots in spin-qubit processors increases in the future, operating them will be even more complex. Qubit routing, then, becomes a key challenge, especially in ever-increasing topology sizes.

Turning now our attention to other qubit technologies, the central distinguishing aspect of spin-qubit quantum processors is the primary communication method used for the routing process. As mentioned, in superconducting devices, qubits are ``moved" employing SWAP gates, in which the quantum state of two neighboring qubits is exchanged. Furthermore, in many superconducting processors, the absence of a native SWAP operation necessitates decomposition in multiple CNOT gates for its implementation, a process that can incur significant errors. In contrast, in spin-qubit architectures, although SWAP operations are supported as well, shuttling is the preferable communication operation as it offers substantial advantages, including higher operational fidelity and quicker execution time. Moreover, certain quantum dots are deliberately left unoccupied to create free space for facilitating qubit movements, as the shuttle process entails the physical relocation of a qubit to an adjacent vacant quantum dot.


However, this introduces new challenges, especially in the context of a quantum compiler routing process, as highlighted in \cite{SpinQ}. This is because, at any given moment, there can be only one qubit per site; therefore, it is not physically possible to shuttle a qubit in an already occupied quantum dot. Consequently, there is a need to develop new qubit routing strategies that avoid such conflicts and efficiently move obstacle qubits away. Other limitations are related to the unique constraints imposed by the classical control electronics, especially in those with shared control schemes \cite{SpinQ}, which might create conflicts in certain cases when trying to apply parallelized shuttle operations. 

Broadly speaking, the main objective of a routing algorithm is to determine the most efficient path(s) to satisfy a list of, either parallelized or not, two-qubit gates while respecting the architecture's constraints and avoiding conflicts. Starting with a defined topology, initial physical and virtual-to-physical qubit placement, the algorithm receives a sequence of two-qubit gates, which, in most cases, cannot be executed directly. Therefore, the primary objective of a router is to identify and correctly integrate qubit movement operations (i.e., shuttles or SWAPs) into the existing gate sequence to ensure that all two-qubit gates can be executed. The overarching goal is to minimize the additional circuit overhead required to achieve this. In the end, the final revised sequence, which includes these movement operations, constitutes the output of our routing algorithm. This process was shown to be in complexity class NP-complete \cite{siraichi2018qubit,botea2018complexity} and thus necessitates a fast solution, as the goal is to use it for scalable architectures that will support thousands of qubits in all sorts of topologies. To date, no algorithmic solution has been proposed to tackle this specific path-finding problem in a practically efficient way for the plethora of unique characteristics of spin-qubit architectures.

\section{Related work and its limitations} \label{Related work}

\begin{table*}[htpb]
\centering
\caption{Comparison of beSnake and Shuttle-based SWAP.}

\includegraphics[width=1\textwidth]{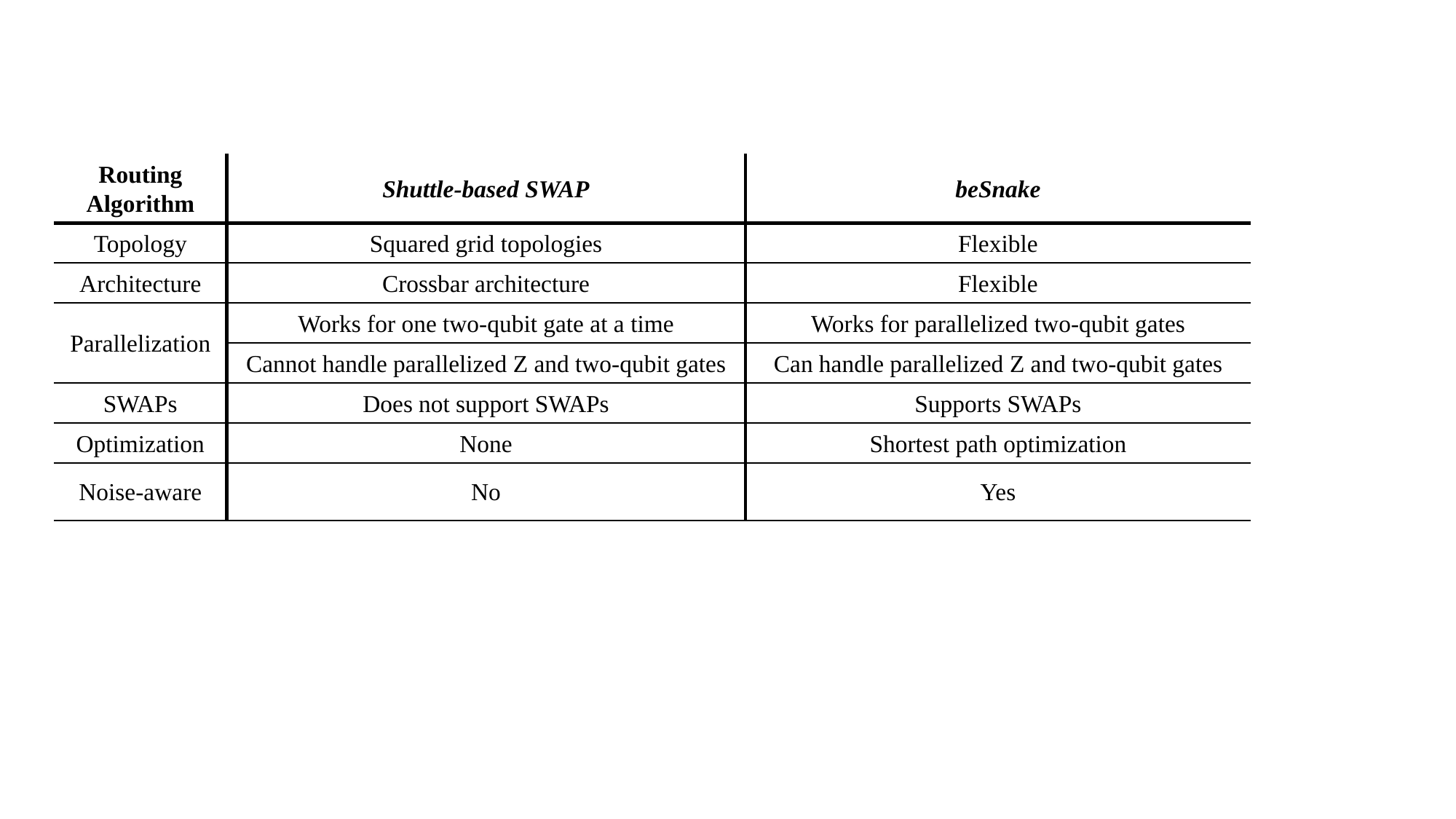}
    \label{fig:comparison_table}
\end{table*}

One can find problems in classical computer science that are similar to qubit routing in spin-qubit architectures. One example is the sliding tile puzzle (STP) \cite{hordern1986sliding} that aims to achieve a specified arrangement with the least possible moves within a squared grid by orthogonally sliding tiles into an empty slot. One established method for addressing this problem has been the Interactive Deepening A* (IDA*) algorithm assisted with a precomputed pattern database to increase the accuracy of the admissible heuristic \cite{korf2002disjoint,felner2004additive}. In more recent work, a 24-tile STP instance was solved using the  Levin Tree Search with Context Models (LTS-CM) \cite{orseau2023levin}. This machine-learning algorithm learned from $50,000$ solved instances of the puzzle and surpasses previous approaches with fewer number of steps. Although these two techniques and others have been very effective, they are not a good candidate for the qubit routing problem for scalable spin-qubit devices: (a) these solutions heavily rely on fine-tuned datasets of particular instances and arrangements of the puzzle. In our problem, creating such a dataset will require large computation resources as we are not aiming for one instance or one specific tile (qubit) configuration but rather a list of goals (i.e., making pairs of qubits adjacent). (b) Another factor to consider with point (a) is that with more moving directions per location (higher node degree), the tree of possible moves expands exponentially in size \cite{gozon2023computing}, as well. The presented STP results in the literature usually assume these datasets or the training phase and do not disclose the total computational costs. However, upon taking a closer look, it becomes obvious that as the number of qubits increases, these approaches become practically intractable. Our problem requires an algorithm that prioritizes speed before optimality (regarding the number of routing steps) to quickly satisfy multiple goals for more than 24 moving parts. (c) As previously mentioned, during circuit executions there are many steps, some of which involve parallelized two-qubit gates. This contrasts with the objective of aiming for a single tile pattern in the STP problem.

Another potentially fitting formulation is the multi-agent path-finding (MAPF) problem, which has been studied extensively for multi-robot systems \cite{gao2023review}. Each agent, representing an independent robotic system, navigates to a target destination through various techniques while avoiding collisions with other agents or obstacles. Agents are sequentially executing a plan consisting of specific goal destinations while trying to minimize the traveled distance within a goal. We have identified, however, the following key differences to our problem: (a) As shown in \cite{gao2023review}, classical MAPF approaches become slow and unsuccessful for large numbers of agents, especially with the existence of obstacles. (b) Our problem is not limited to a single set of goals (i.e., a set of two-qubit gates) but a plan with multiple goals in each step. Therefore, ultimately, the focus is on minimizing some cost function for the entire plan (representing a quantum circuit), as it is crucial to incur the least errors possible. (c) In our problem, qubits can be viewed as agents and as dynamic obstacles at the same time. In addition, only a subset of agents (qubits) need to complete goals at each step of the plan, and these goals are not target locations but conditions that could be satisfied in multiple locations. Currently, there is a lack of strategies specifically tailored to address this particular problem in a short amount of time, especially when dealing with a large number of agents, as highlighted in \cite{raja2012optimal,ma2022graph,tjiharjadi2022systematic}.

In the quantum world, several qubit routing techniques have been proposed, but mostly for superconducting and trapped ion platforms as they are the most advanced in terms of qubit counts. Routing for spin qubits resembles a similar functionality of quantum charge-coupled ion trap devices (QCCD) \cite{bruzewicz2019trapped}, another promising scalable architecture. In QCCD, trap regions are dedicated to specific tasks such as temporary storage or processing. Communication between ions of different regions is thus implemented with shuttles through a shared channel where collision avoidance optimization techniques are used \cite{bruzewicz2019trapped, murali2020architecting}. However, these techniques fundamentally differ because they are made specifically for the QCCD's unique regions and topology. The unique structure of a QCCD device imposes a distinct set of constraints on moving qubits and parallelizing operations, different from to those found in 2D spin qubit topologies. Hence, these routing strategies cannot be implemented in spin qubit platforms even though both use shuttling \cite{saki2022muzzle,schoenberger2023using}.

In superconducting systems, routing involves consecutive state exchanges by means of SWAP gates along a chosen path until the desired qubit pair is adjacent. The most commonly used methods for qubit routing in such devices are based on heuristic search algorithms or reinforcement learning techniques \cite{murali2020architecting,saki2022muzzle,schmale2022backend,schoenberger2023using,pozzi2022using}], although recently more theoretical proposals have been introduced for bounding the minimal number of SWAPs needed \cite{steinberg2024resource,escofet2024revisiting}. The only requirement to determine a viable path is the presence of at least one pair of neighboring locations within the topology where the two qubits can interact. In spin-qubit architectures, on the other hand, routing entails the physical relocation of qubits instead of state swapping. Therefore, within the notion of bringing two qubits closer to each other until adjacent, other qubit positions play an important role as well.

When routing in superconducting devices, the predominant approach is to identify the shortest paths between qubit operands, concurrently minimizing costs across several parameters. These parameters include circuit gate and depth overhead, as well as noise-aware path selection \cite{murali2020architecting,saki2022muzzle,schmale2022backend,schoenberger2023using,pozzi2022using}. These can be optimization goals for a spin-qubit routing algorithm as well, but finding the shortest path is not enough because other qubits might block the way. The problem can become even more complex when considering routing for multiple two-qubit gates at the same time while trying to avoid conflicts and satisfy operational constraints of the architecture \cite{SpinQ}.
 

The Shuttle-based SWAP (SBS) algorithm was conceptualized as the first routing algorithm for large-scale spin qubit crossbar architectures \cite{SpinQ}. In particular, it was tailored to the unique and complex constraints of this crossbar architecture \cite{li2018crossbar}, which necessitated the maintenance of the checkerboard pattern of physical qubits to achieve a fast compilation process. One of the main drawbacks of these constraints was the parallelization limitations imposed on shuttle operations. As a consequence, SBS can only, yet efficiently, route one two-qubit gate at a time on grid topologies for that particular crossbar architecture. As extensively discussed in \cite{SpinQ}, this severely impacts the mapping overhead for high qubit counts and high two-qubit gate percentages. 

In this paper, we are extending the work on spin qubit routing to fully utilize and explore the benefits of shuttling-based movement of qubits, assuming a more generalized architecture with fewer constraints. We propose beSnake, a qubit routing algorithm for large-scale spin qubit processors, which can handle any combination and number of parallelized shuttle-based Z and two-qubit gates in any topology. Furthermore, beSnake is designed to be more flexible, which opens opportunities for more complex routing tasks. We are highlighting their main differences in Table \ref{fig:comparison_table}. Based on the early stage of spin-qubit technologies, beSnake is introduced as a concrete baseline routing algorithm for future scalable architectures that use shuttling.

\section{The beSnake routing algorithm} \label{The beSnake routing algorithm}

The main objective of beSnake is to enable non-adjacent operand qubits of two-qubit gates interact with each other by strategically shuttling qubits around the topology. It does so by considering multiple shortest paths between the operands and subsequently implementing an adjustable path selection heuristic to potentially minimize the added movements as much as possible. Upon a path selection, beSnake uses a breadth-first search (BFS) exploration for shuttling qubits; hence, the first part of beSnake's name comes from the first letter of BFS. This approach with the particular shortest path selection ensures the minimization of shuttle operations, viewed from the operand's perspective, significantly reducing the pool of qubits considered for movement. In terms of managing parallelized gates, each is routed in turn, while their operands are held in place once they become adjacent until all of the gates are attempted. In case of failed attempts, three strategies are implemented to overcome them, discussed in Section \ref{Mechanisms for blockades}, including a lookback mechanism. In the end, the gates can be simultaneously executed, thereby maintaining the circuit depth as low as possible.  
 
In order to further elaborate on the functionalities included in beSnake, we will use three representative cases based on the squared grid topology shown in Fig. \ref{fig:topologya}. In these examples, we will use the notion of the intermediate representation (IR) for the input circuits as well as output routed circuits with the variable names \texttt{ir\_input} and \texttt{ir\_output}, respectively. Within the IR, the gates of a cycle are placed inside square brackets, and commas separate each gate. Note that the notion of a cycle refers to the basic unit of time representing one step in a sequence of gates of a quantum circuit, and each step may contain multiple gates. Moving on, the temporary \texttt{goals} variable contains a single cycle of the input circuit given to beSnake for routing, and the \texttt{goal} variable holds the current gate from \texttt{goals} to be routed. Finally, we will use the \texttt{prev\_goal\_qubits} variable, which contains the pairs of operand qubits that have already been satisfied from \texttt{goals}. 

\subsection{Case 1: Following a free path}

\begin{figure*}[htpb]
     \centering
     \begin{subfigure}[b]{0.17\textwidth}
         \centering
         \includegraphics[width=\textwidth]{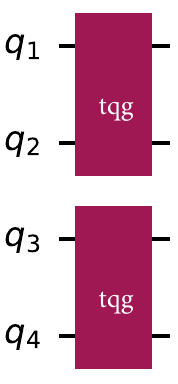}
         \caption{ }
         \label{fig:circuit}
     \end{subfigure}
\hfill
     \begin{subfigure}[b]{0.497\textwidth}
         \centering
         \includegraphics[width=\textwidth]{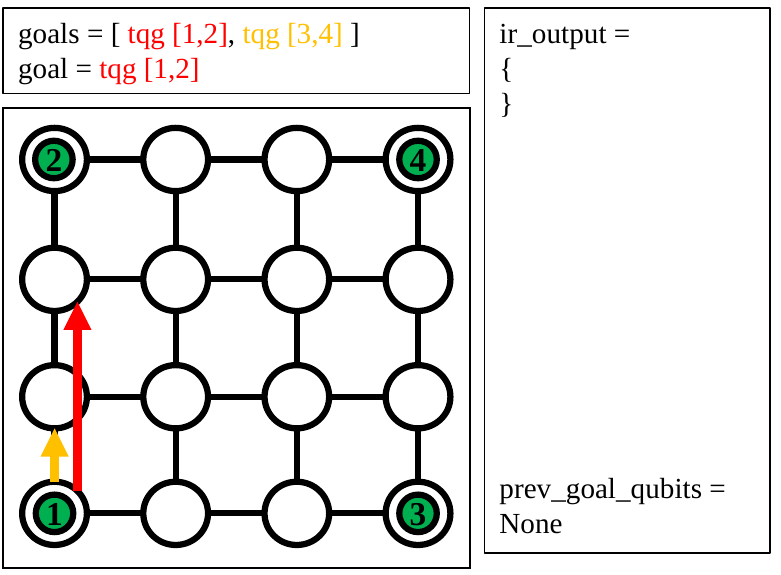}
         \caption{ }
         \label{fig:firstexamplea}
     \end{subfigure}
     \begin{subfigure}[b]{0.497\textwidth}
         \centering
         \includegraphics[width=\textwidth]{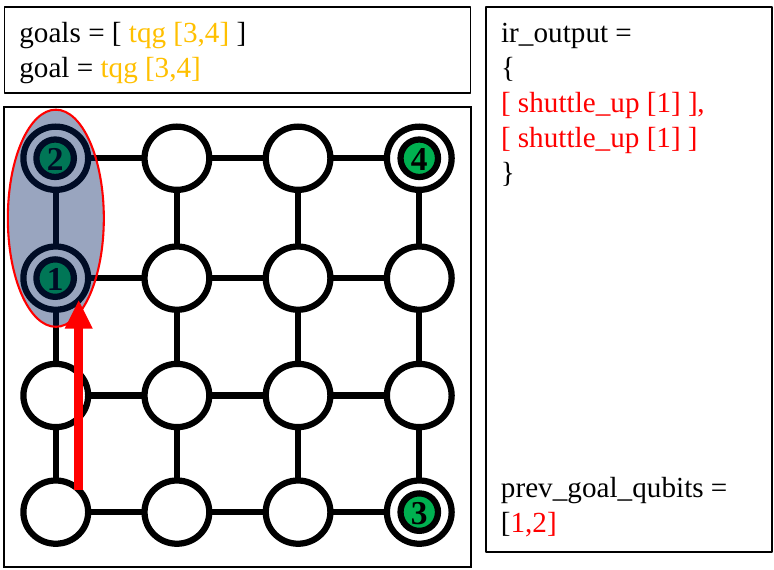}
         \caption{ }
         \label{fig:firstexampleb}
     \end{subfigure}
     \begin{subfigure}[b]{0.497\textwidth}
         \centering
         \includegraphics[width=\textwidth]{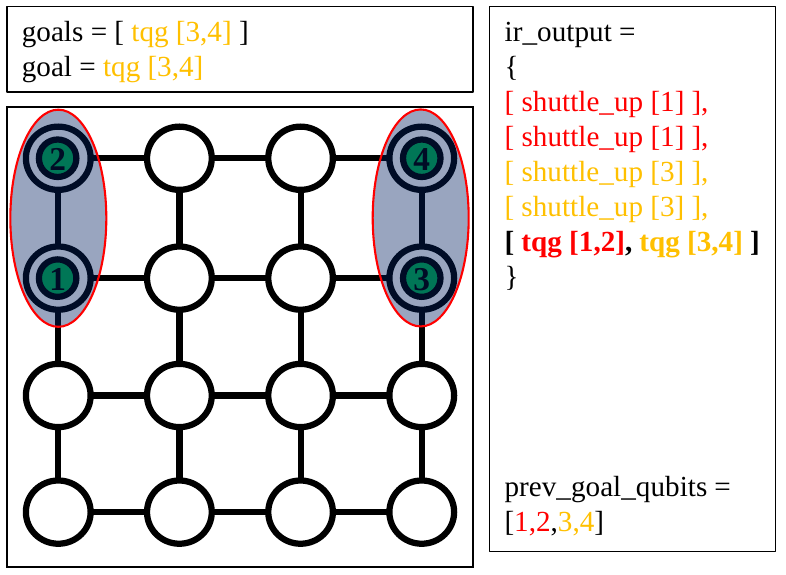}
         \caption{ }
         \label{fig:firstexamplec}
     \end{subfigure}
        \caption{Example showcasing how beSnake follows the shortest path to enable two-qubit gate interactions. \textbf{(a)} Diagram circuit representation given to beSnake for routing with the following IR input \texttt{ir\_input = \{ [ tqg [1,2], tqg [3,4] ] \}} \textbf{(b)} The first step is to consider the shortest paths of the first gate to be routed. In this case, there is only one, shown with the red arrow. The yellow arrow represents the first step of qubit 1 following the shortest path. \textbf{(c)} The completion of the first \texttt{goal} with the two qubits becoming adjacent and fixed until the rest of the gates in variable \texttt{goals} are tried. \textbf{(d)} The completion of routing for all gates in \texttt{goals} and the final IR output.}
        \label{fig:firstexample}       
\end{figure*}

The first example illustrates the most simple case of qubit routing. We assume a quantum circuit with only two two-qubit gates (\texttt{tqg}) in the first cycle (see Fig. \ref{fig:circuit}), whose interacting qubits are placed in non-neighboring positions (see Fig. \ref{fig:firstexamplea}) and therefore require routing. The corresponding IR representation of the quantum circuit can be derived as \texttt{ir\_input = \{ [ tqg [1,2], tqg [3,4] ] \}}. Note that there can be more than one cycle in an IR, but we assume one for illustrative purposes. As depicted in Fig. \ref{fig:firstexamplea}, the beSnake algorithm takes the first gate of the cycle (variable \texttt{goals}) and assigns it to \texttt{goal}. It then calculates all possible shortest paths between its operands (1 and 2). In this case, there is only one shortest path from qubit 1 to qubit 2, comprising just two nodes. More precisely, qubit 1 will need to move upwards two times.

After these movements, qubits 1 and 2 will be adjacent, as shown in Fig. \ref{fig:firstexampleb}, and they will remain in that position until the two-qubit gate is executed. Note that the \texttt{ir\_output} list is updated accordingly, including two new shuttles. In addition, the algorithm updates \texttt{prev\_goal\_qubits} with qubits 1 and 2, indicating that they will remain fixed in their current positions until the rest of the gates are routed. It then proceeds to the next gate in \texttt{goals} and repeats the same process. The outcome, as shown in Fig. \ref{fig:firstexamplec}, is that the qubits of both gates within \texttt{goals} have been successfully routed, and now the \texttt{ir\_output} encompasses all necessary shuttles to position the qubit operands adjacent to each other. After that, the two-qubit gates can be executed together as originally given from the \texttt{ir\_input}. At this stage, beSnake has completed the task, and more optimization passes can further improve the output \texttt{ir\_output}, such as a scheduler parallelizing even more the shuttles required for moving qubits 1 and 3.

This example showed a straightforward operation; however, as we will explore in subsequent examples, certain qubits may block the routing path and, therefore, necessitate relocation to open the way.

\subsection{Case 2: Following a blocked path}

\begin{figure*}[htpb]
     \centering
     \begin{subfigure}[b]{0.497\textwidth}
         \centering
         \includegraphics[width=\textwidth]{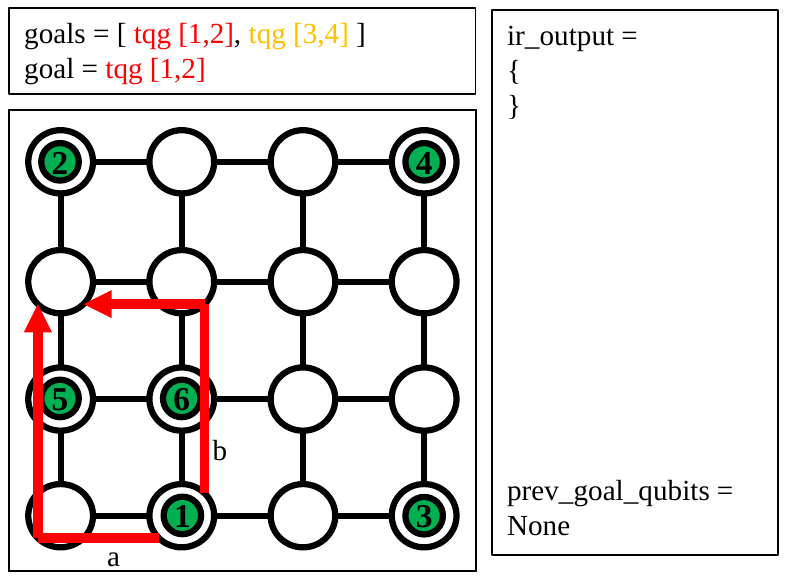}
         \caption{ }
         \label{fig:secondexamplea}
     \end{subfigure}
     \begin{subfigure}[b]{0.497\textwidth}
         \centering
         \includegraphics[width=\textwidth]{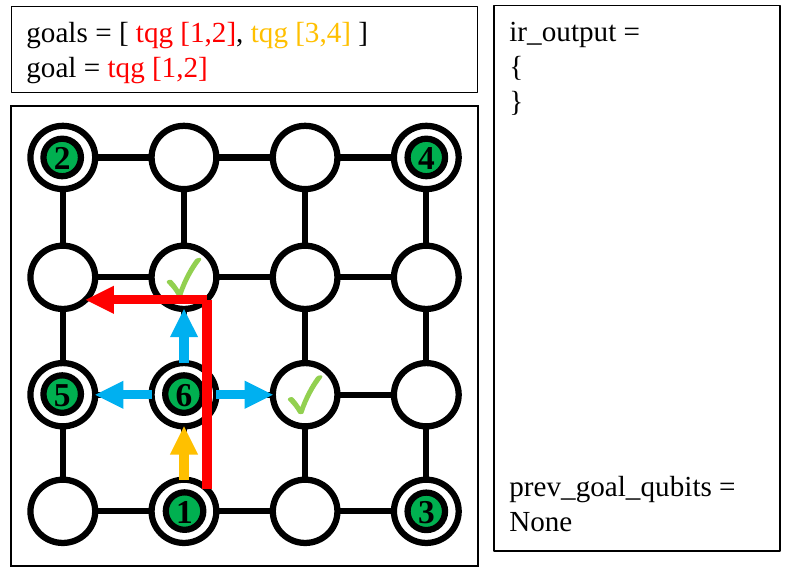}
         \caption{ }
         \label{fig:secondexampleb}
     \end{subfigure}
     \begin{subfigure}[b]{0.497\textwidth}
         \centering
         \includegraphics[width=\textwidth]{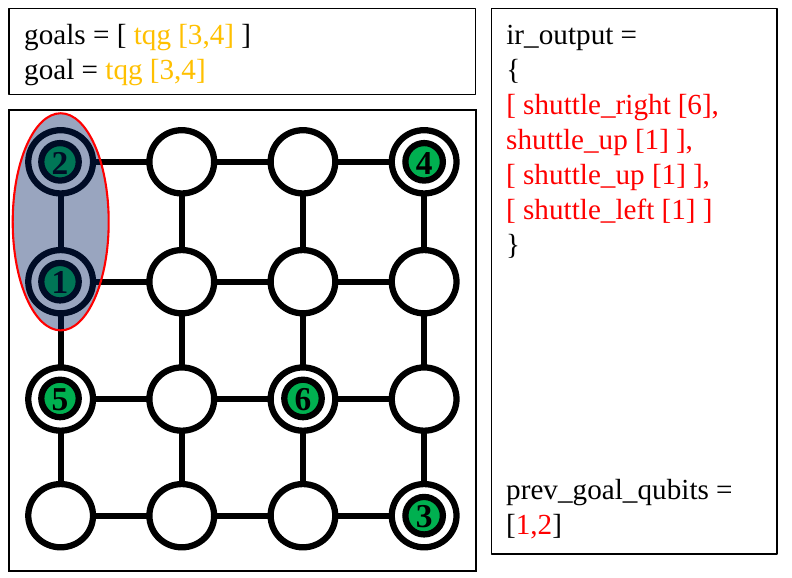}
         \caption{ }
         \label{fig:secondexamplec}
     \end{subfigure}
        \caption{Example showcasing how beSnake resolves routing scenarios with blockading qubits within the shortest path for a circuit with \texttt{ir\_input = \{ [ tqg [1,2], tqg [3,4] ] \}}. \textbf{(a)} The first step is to consider the shortest paths and select one based on two heuristic criteria. For illustration purposes, two paths are drawn with an equal number of qubit obstacles. However, path \texttt{b} traverses a path with higher accumulated degrees of the traversed nodes, hence its selection. \textbf{(b)} The BFS exploration of beSnake to move obstacle qubit 6 with the least possible number of shuttles. Each layer of exploration is depicted with different colored arrows until an empty position can be found. \textbf{(c)} The state of the qubit positions after completing the first \texttt{goal}. The example can continue from this point on similarly to the example in Fig. \ref{fig:firstexample}. }
        \label{fig:secondexample}       
\end{figure*}

In our second example, we address a similar situation, but this time, qubits are blocking the shortest paths, as shown in Fig. \ref{fig:secondexamplea}. We can identify four potential shortest paths for addressing the first two-qubit gate between qubits 1 and 2, with each path comprising three steps. For simplicity, we focus only on two paths, labeled \texttt{a} and \texttt{b}, as illustrated.

beSnake employs a two-tiered filtering criterion to optimize the path selection. Firstly, it will keep the path(s) with the least obstacle qubits, and out of the remaining paths, it will select one with the highest accumulated degree of traversed nodes. This is to mitigate the likelihood of congestion in the less connected areas, such as the corners of the current topology. This heuristic aims to minimize obstacle encounters and to reduce the overhead associated with qubit movement.

Upon comparison, both paths \texttt{a} and \texttt{b} involve a single obstacle qubit; however, path \texttt{b} has notably higher accumulated degree of the traversed nodes, satisfying our selection criteria. When initiating the first step of the preferred shortest path \texttt{b}, as indicated by an orange arrow in Fig. \ref{fig:secondexampleb}, we encounter a blockage due to qubit 6. Then, an exploration of potential movements for qubit 6 to clear the path is conducted and three possible directions are identified: moving left towards qubit 5, moving up along the shortest path, and moving to the right. Of these possible movements, only one is practical -- moving to the right. By doing so, we avoid the additional overhead that would otherwise arise later by moving it again away from the shortest path.

Fig. \ref{fig:secondexamplec} showcases qubit 1 having reached qubit 2, updating \texttt{prev\_goal\_qubits} and removing the gate from \texttt{goals}. The progression from this point on is similar to the one presented in Fig. \ref{fig:firstexamplec}.

\subsection{Case 3: Satisfying mutiple two-qubit gates} \label{Third Example: Satisfying parallelized gates}

\begin{figure*}[!ht]
     \centering
     \begin{subfigure}[b]{0.497\textwidth}
         \centering
         \includegraphics[width=\textwidth]{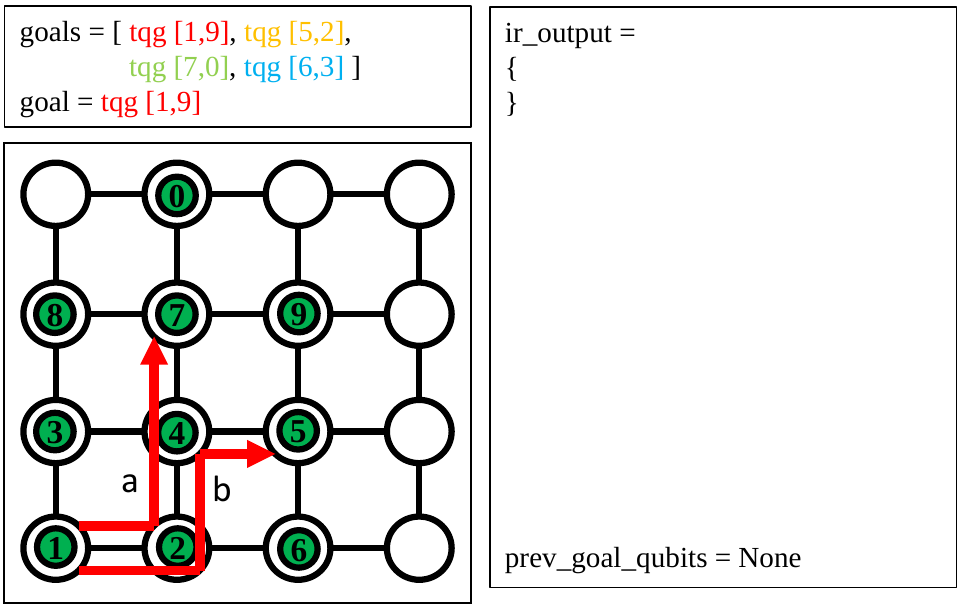}
         \caption{ }
         \label{fig:thirdexamplea}
     \end{subfigure}
     \begin{subfigure}[b]{0.497\textwidth}
         \centering
         \includegraphics[width=\textwidth]{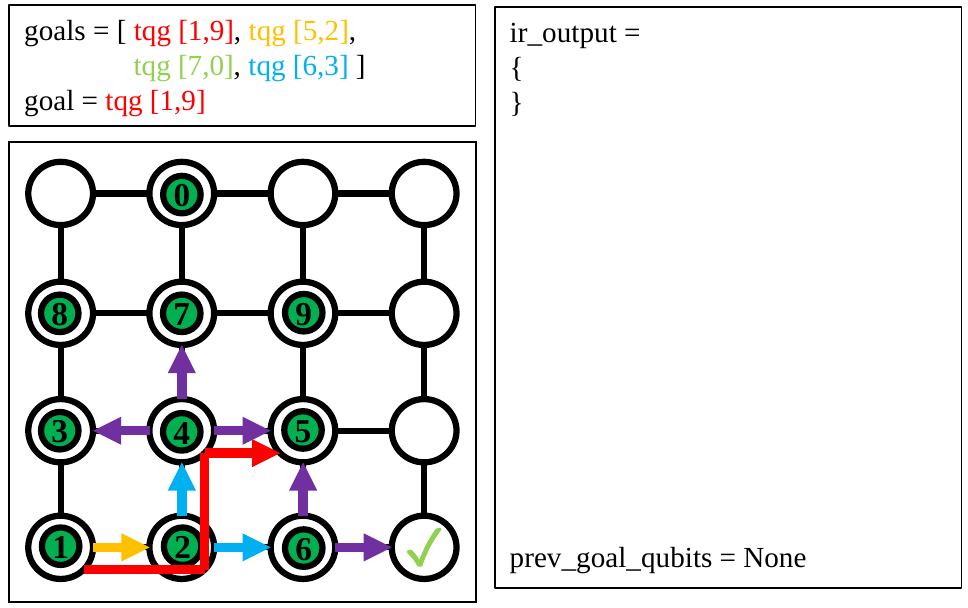}
         \caption{ }
         \label{fig:thirdexampleb}
     \end{subfigure}
     \begin{subfigure}[b]{0.497\textwidth}
         \centering
         \includegraphics[width=\textwidth]{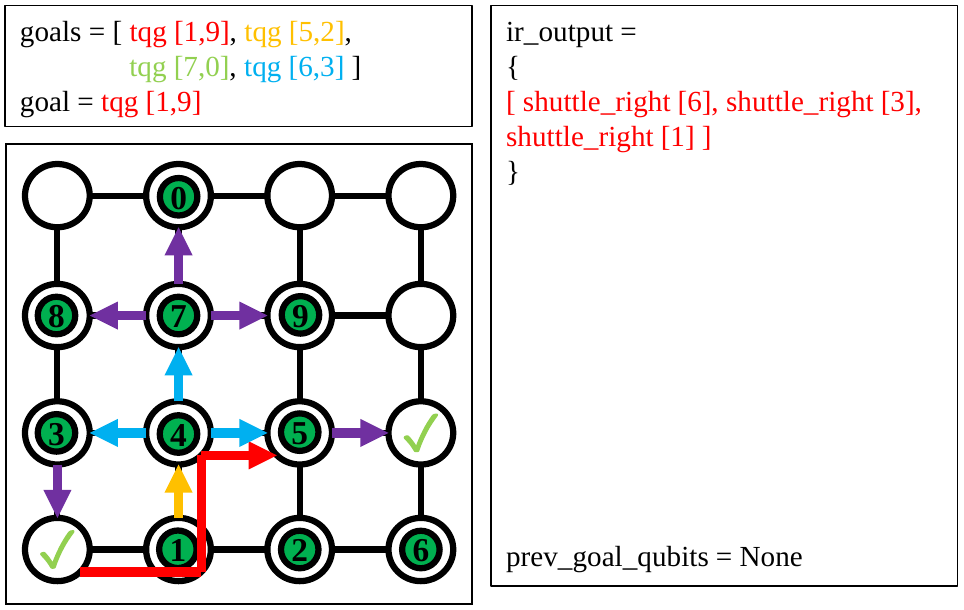}
         \caption{ }
         \label{fig:thirdexamplec}
     \end{subfigure}
     \begin{subfigure}[b]{0.497\textwidth}
         \centering
         \includegraphics[width=\textwidth]{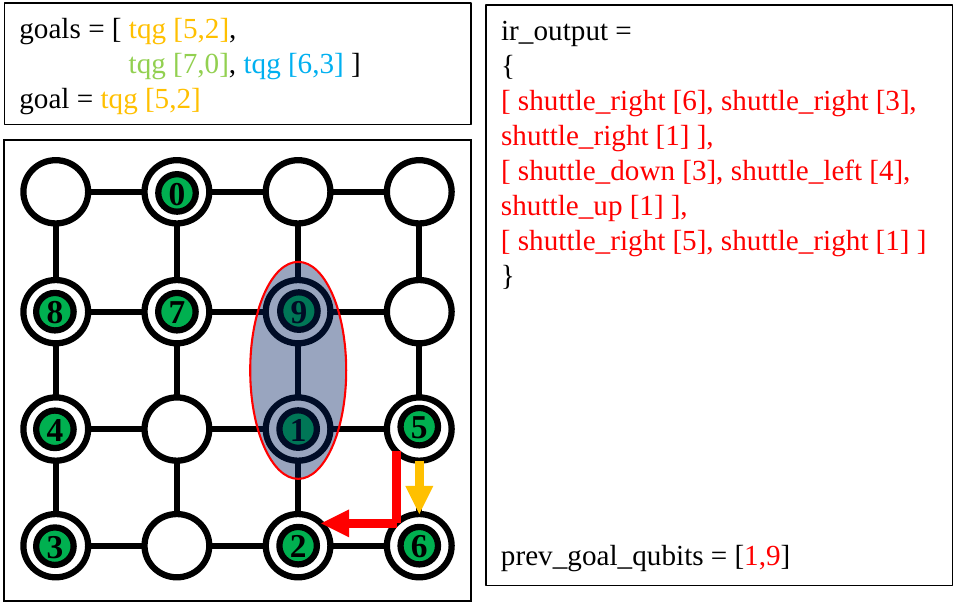}
         \caption{ }
         \label{fig:thirdexampled}
     \end{subfigure}  
     \begin{subfigure}[b]{0.497\textwidth}
         \centering
         \includegraphics[width=\textwidth]{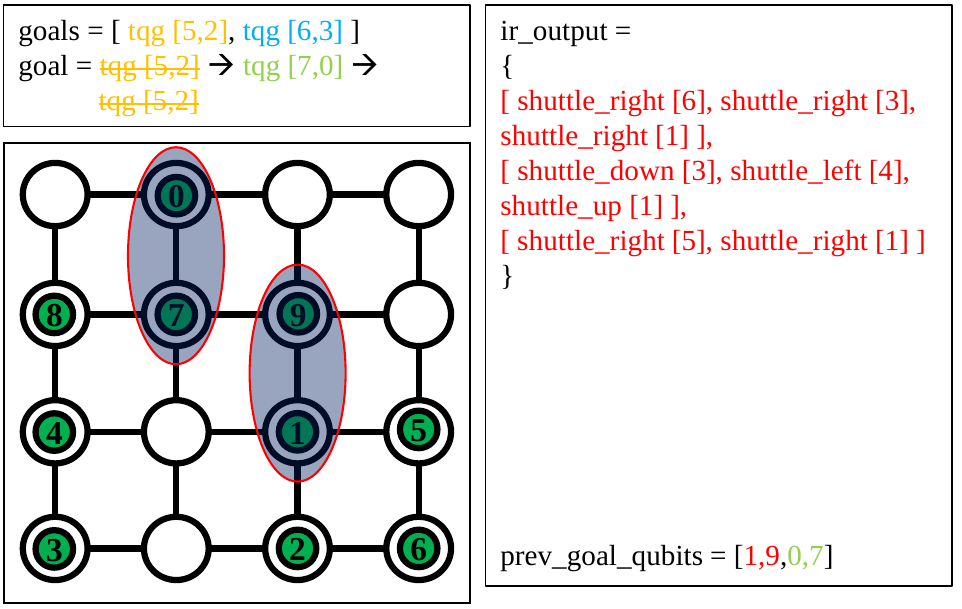}
         \caption{ }
         \label{fig:thirdexamplee}
     \end{subfigure}  
     \begin{subfigure}[b]{0.497\textwidth}
         \centering
         \includegraphics[width=\textwidth]{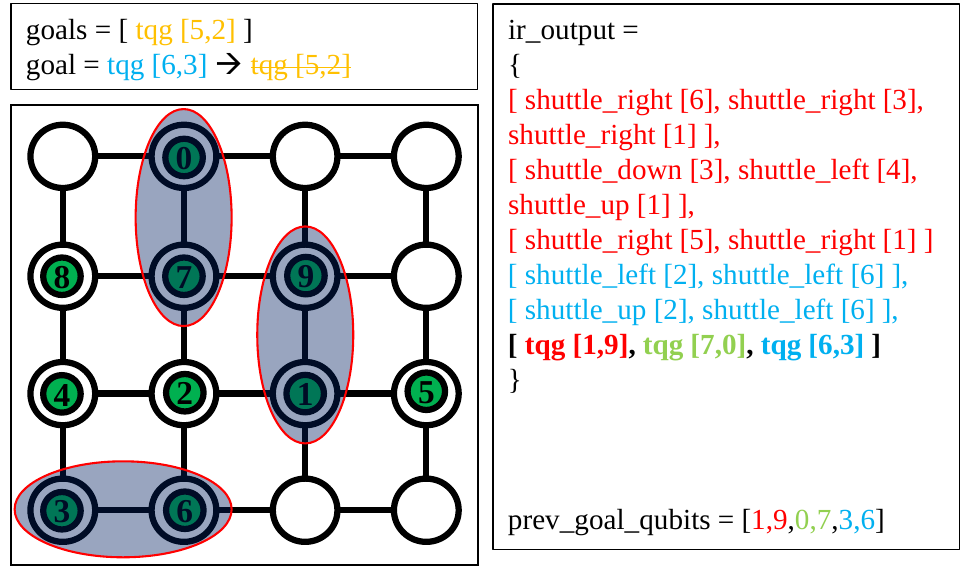}
         \caption{ }
         \label{fig:thirdexamplef}
     \end{subfigure}     
        \caption{Example showcasing how beSnake resolves more complex routing scenarios with multiple blockading qubits for a circuit with \texttt{ir\_input = \{ [ tqg [1,9], tqg [5,2], tqg [7,0], tqg [6,3] ] \}}. \textbf{(a)} For illustration purposes, we present only two shortest paths that equally satisfy our two heuristic criteria. Path b is selected for this example. \textbf{(b)} The BFS exploration of beSnake to move obstacle qubit 2 away with the least number of shuttles. \textbf{(c)} The next BFS exploration for moving obstacle qubit 4 away with the least number of shuttles. Two solutions are found in this case, and beSnake selects one based on the least occurred error rate. \textbf{(d)} The completion of the first gate in \texttt{goals} and the exploration of shortest paths for the next, which can not be resolved at this stage. \textbf{(e)} The resolution of the next available gate in \texttt{goals} and the revisiting of the previous failed one. \textbf{(f)} Completion of the next available gate in \texttt{goals} and the scheduling of all so-far successful \texttt{goals} in \texttt{ir\_output}. (Continues on the next page)}
        \label{fig:thirdexample}       
\end{figure*}
     
\begin{figure*}[!ht]
\ContinuedFloat 
\centering
\begin{subfigure}{0.497\textwidth}
  \centering
 \includegraphics[width=\textwidth]{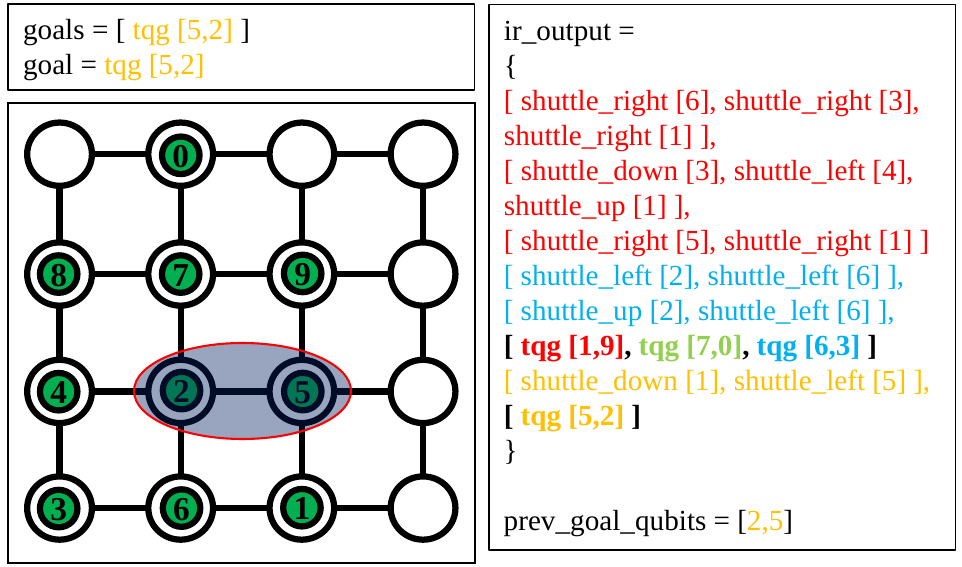}  \caption{ }
 \label{fig:thirdexampleg}
\end{subfigure}%
\caption{ (Fig. \ref{fig:thirdexample} continuation) \textbf{(g)} The previously fixed positions of the scheduled two-qubit gates are freed, and now the two-qubit gate \texttt{tqg [5,2]} can finally be routed.}
\end{figure*}

In the complex scenario depicted in Fig. \ref{fig:thirdexamplea}, beSnake is tasked with satisfying four gates within a cycle amidst a denser qubit topology. By considering routing for these four gates simultaneously, we can achieve a lower depth overhead compared to routing for each separately. We focus on the initial gate listed in \texttt{goals}. As always, we look for the shortest path with the least number of qubit obstacles and with the largest number of adjacent edges. Among the six conceivable paths, each shares an identical qubit count; however, only two paths have the most connections to other vertices. The algorithm resolves ties by randomly selecting a path, path \texttt{b} for this example.

The subsequent phase involves determining the sequence of shuttle movements required to reposition qubit 1 one step further in the chosen shortest path. Fig. \ref{fig:thirdexampleb} illustrates the exploratory moves beginning from qubit 1. Intuitively, we can envision this as a BFS style of exploration where qubits sequentially push each other until the first vacancy is encountered. Expressly, qubit 1 is set to push qubit 2 (indicated by an orange arrow), while qubit 2 tries to push qubits 4 and 6 (blue arrows), and qubits 4 and 6 try to push qubits 3, 5, and 7 (purple arrows). At this point, a solution has been identified, rendering additional exploration unnecessary, as it would certainly increase the number of moves.

Fig. \ref{fig:thirdexamplec} updates \texttt{ir\_output} with the first sequence of moves from the first step of qubit 1, within a single cycle. This figure also displays the exploration of qubit 1's next move, with each expansion layer marked by arrows colored orange, blue, and purple, respectively. At this point, we see two possible solutions, moving qubits 3 and 5. Both solutions need the same number of shuttles. beSnake then becomes noise-aware by selecting the one with the highest accumulated shuttle fidelity based on predefined fidelity attributes for each coupling connection. The accumulated fidelity is calculated by multiplying the fidelities of each shuttle on the specific locations. This mechanism is used in case different fidelities are associated with each coupling link, thus maximizing the routed circuit's fidelity. As a result, the movement of qubit 3 is chosen for this example, and in Fig. \ref{fig:thirdexampled}, the sequence of shuttles is displayed in the second cycle of \texttt{ir\_output}. In the same figure, we have fast forward to the final step of the shortest path and see the three cycles within \texttt{ir\_output} and updates to \texttt{prev\_goal\_qubits}. The \texttt{goal} now includes the next gate from \texttt{goals} and the previous gate has been removed. A single shortest path is discovered since it is not allowed to push fixed qubits, yet qubit 6 is immobilized. This is because the operand qubits are prohibited from being pushed by any other qubit as well. Consequently, beSnake omits this gate and proceeds to the subsequent one. This mechanism is one of three mechanisms implemented (see section \ref{Mechanisms for blockades}) to deal with unsuccessful attempts when routing certain two-qubit gates. In this particular case, we will move on to route the next available gate in \texttt{goals}, the \texttt{tqg [0,7]}, which might potentially move qubits around the topology and free the blockage shown in Fig. \ref{fig:thirdexampled}.

Fig. \ref{fig:thirdexamplee} shows that the next goal is immediately satisfied, and the algorithm reattempts the routing of the previously unsuccessful gate, the \texttt{tqg [5,2]}. This is the lookback mechanism used to iterate over the remaining gates in \texttt{goals}. Once \texttt{tqg [7,0]} is accomplished, it is removed from \texttt{goals}, and the algorithm revisits the \texttt{goals} set, starting with any gates that previously failed. However, the \texttt{tqg [5,2]} remains unsuccessful due to the same reasons. Fig. \ref{fig:thirdexamplef} demonstrates the resolution of the following gate, \texttt{tqg [6,3]}, the next gate in \texttt{goals}, at which stage all gates have been satisfied except for one. A repeated attempt yields no solution, as no shortest path is viable anymore between qubits 5 and 2. Due to this fact, we update \texttt{ir\_output} by scheduling only the satisfied gates, thus splitting the original cycle.

Subsequently, all qubits in \texttt{prev\_goal\_qubits} are released, and the algorithm reattempts the unsolved gate. This time, a solution is possible, and Fig. \ref{fig:thirdexampleg} depicts the successful outcome.

\subsection{Special cases}

This section introduces the mechanisms employed by beSnake to address special routing cases. We outline three primary mechanisms utilized to circumvent blockades showcasing the algorithm's potential to adapt to conflicts.

\subsubsection{Mechanisms for dealing with blockades} \label{Mechanisms for blockades}

An obstacle qubit might disallow the movement of any operand qubit within the shortest path. For example, in Fig. \ref{fig:thirdexamplec}, obstacle qubit 6 is blocking the very first step in the shortest path. However, there can be cases where a similar situation is reached at any step in the path due to qubits being pushed in areas of the topology with fewer connections, such as the corners of squared topologies. Then, the following mechanisms will be tried in this order:
\begin{enumerate}
    \item Revisit the same gate (\texttt{goal}) after satisfying other ones. This lookback mechanism was used in Fig. \ref{fig:thirdexamplee}. 
    \item Splitting the original cycle and trying the remaining gate(s) without the previously fixed qubits. This mechanism was used in Figures \ref{fig:thirdexamplef} and \ref{fig:thirdexampleg}. 
    \item Execute a forced SWAP between the obstacle and the operand qubit.
\end{enumerate}

Since the first two mechanisms have been demonstrated in Section \ref{Third Example: Satisfying parallelized gates}, we are now focusing on the third one; If we isolate the scenario in Fig. \ref{fig:thirdexampled} and suppose the first two mechanisms failed, then we will insert a forced SWAP between qubit 5 and 2 to overcome this obstacle. This mechanism was implemented to efficiently resolve such rare scenarios where qubit obstacle(s) are located in between both operands while having no available edges to be pushed away. This can be triggered in topologies that may have only two coupling links at one or more nodes (quantum dots). In the case of a square grid, this place is at the corners. Alternatively, this can be addressed by taking another path, not necessarily a shortest path, that avoids such restricted regions. However, this does not guarantee a solution and most likely will result in more overhead. In Section \ref{Conclusions}, more alternatives are discussed.


\subsubsection{Handling shuttle-based Z rotations} \label{Handling shuttle-based Z rotations}

\begin{figure*}[htpb]
     \centering
     \begin{subfigure}[b]{0.26\textwidth}
         \centering
         \includegraphics[width=\textwidth]{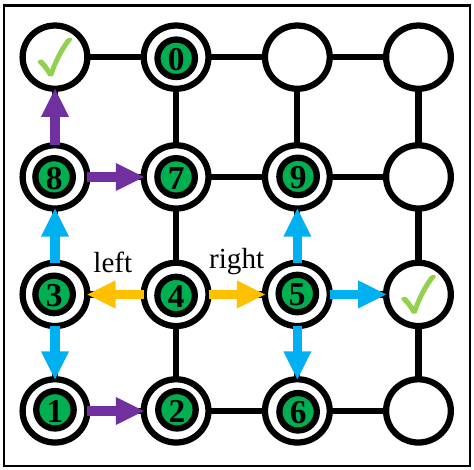}
         \caption{ }
         \label{fig:Za}
     \end{subfigure}
    \begin{subfigure}[b]{0.26\textwidth}
         \centering
         \includegraphics[width=\textwidth]{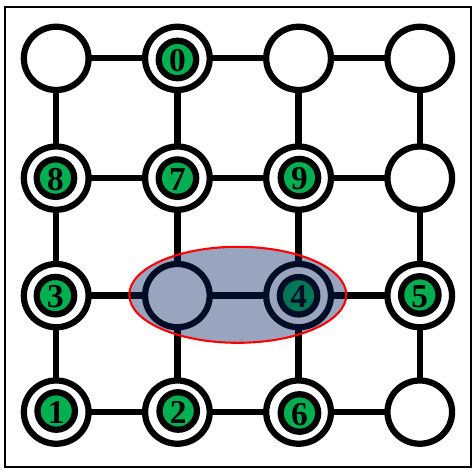}
         \caption{ }
         \label{fig:Zb}
     \end{subfigure}
             \caption{An example of how beSnake optimizes the number of shuttles for shuttle-based Z gates. \textbf{(a)} beSnake will explore both possibilities to shuttle qubit 4 left and right and select the one with the least moves. In this case, moving qubit 5 to the right requires fewer shuttles than trying to move qubit 3. \textbf{(b)} The qubit positions after the shuttles and the fixation of positions to ensure qubit 4 can return in time. }
        \label{fig:Z}       
\end{figure*}

Shuttle-based Z rotation gates are a unique high-fidelity implementation of a Z rotation with two time-sensitive qubit shuttles to and from a neighboring column in either direction \cite{SpinQ,helsen2018quantum,morais2019mapping,li2018crossbar}. This kind of single-qubit manipulation technique, otherwise called ``hopping spins" \cite{wang2024operating}, can, in fact, achieve an arbitrary rotation axis, but in this work, we assume the proposal in \cite{li2018crossbar} for simplicity. beSnake can determine which direction (left or right) will result in the least sequence of shuttles. Then, it will fix the positions (one empty and one containing the qubit operand of the Z gate) so that the return shuttle can be inserted immediately. To demonstrate that, we suppose a shuttle-based Z rotation on qubit 4 in the example of Fig. \ref{fig:thirdexamplea}. This will result in the example shown in Fig. \ref{fig:Za}, where moving to the right takes fewer moves than moving left. In Fig. \ref{fig:Zb}, the original (now empty) location of qubit 4 is not allowed to change until the return shuttle is executed. More specifically, if the empty location gets occupied before the return shuttle, qubit 4 cannot return in time. In case there are parallel Z gates to be satisfied in a cycle, beSnake will follow the same logic as described in Figures \ref{fig:thirdexamplee} and \ref{fig:thirdexamplef}, and iteratively try all of them. An exception is applied here where succeeding Z gates can override fixed empty locations from preceding Z gates. Since implementing Z gates involves a second shuttle back to the original location, multiple of these shuttles can take place simultaneously to complete the cycle even if they previously occupied fixed empty locations of other Z gates.  One example of this exception is when a shuttle-based Z gate for qubit 3 is scheduled together with the Z gate on qubit 4 in Fig. \ref{fig:Zb}. Qubit 3 will shuttle to the right since this is the only available direction, overriding the fixed position created by the Z gate of qubit 4. Then, supposing the cycle finishes here, both qubits can return to their original locations with the well-timed return shuttles, completing their Z rotations.  This exception provides better parallelization as extra room is provided for more Z gates while maintaining the same gate overhead (as we would have without this exception). 

Finally, in case there are Z and two-qubit gates in a cycle, the two-qubit gates will be sorted at the beginning of \texttt{goals} to be tried first. This is because routing for two-qubit gates will more likely require multiple steps to complete, and therefore, in the beginning, it is more important to have as few fixed positions as possible (i.e., more free space to push/move qubits). In Section \ref{Conclusions}, more ideas are discussed related to sorting gates in cycles.

\subsection{Optional functionalities} \label{Optional functionalities}

beSnake has two optimization settings that can potentially improve the performance of both the routed circuit and the speed of the algorithm itself. 

\begin{itemize}
    \item  \textbf{SWAP replacement:}
beSnake can become even more noise-aware by replacing a sequence of shuttles facilitating a step within the shortest path by a SWAP operation. One example is the sequence of shuttles inserted in Fig. \ref{fig:thirdexamplec}. It will do so when the accumulated fidelity of those shuttles exceeds that of the SWAP, particularly in these locations. 

\item \textbf{Shortest-path optimizations:}
As described before, beSnake will assess multiple shortest paths and heuristically pick one that will most likely produce the least amount of gate and depth overhead. However, this ability is by far the most time-consuming task of beSnake as the discovery of all shortest paths, and subsequently, their filtration is iterated several times. To mitigate that, there is an option to place a time limit on the discovery of the shortest paths, thus substantially reducing the time this process takes. Additionally, there is an option to take a single shortest path, which significantly speeds up the process. Later in Section \ref{time limit}, we will deliberately test these options and determine the routing time benefits at the expense of increased circuit overhead.

\end{itemize}

\section{Simulations and Evaluations} \label{Simulations}

With the following simulations, our goal is to test beSnake's different functionalities and create valuable performance insights under various use cases. More specifically, we test beSnake's capabilities with different levels of shortest path optimizations and SWAP replacement options, as well as its behavior on more connected topologies and ranges of qubit densities. Based on these insights, we move on to thoroughly compare it with the SBS algorithm on synthetic and real quantum circuits with up to 1000 qubits. 

\subsection{Experimental Setup} \label{Experimental Setup}

The beSnake and SBS routing algorithms have been incorporated into the SpinQ compilation framework \cite{SpinQ} and executed single-threaded on a 2.4GHz and 768GB memory server running Python 3.6.8. 

Regarding the metrics used to determine their performance, we have used the total computation time of routing for each circuit, which we simply named \textit{routing time}. In the case of routing time comparisons, we provide the ratio (\textit{rTime}) of the two results as: (\textit{results\textsubscript{a}} / \textit{results\textsubscript{b}}). It should be noted that routing times can vary depending on the server's total load and dynamic job scheduling at the time of submission. Acknowledging that, we have taken measures to maintain good time consistency between our simulations by submitting jobs in dedicated resources with no influence from other jobs. 

As for overhead metrics, we have used the same gate and depth overhead definitions as in \cite{SpinQ}. \textit{Gate overhead} is calculated as the percentage relation of additional gates incorporated by the router divided by the number of gates (after decomposition). \textit{Depth overhead} is expressed as the percentage of extra depth generated by the router in relation to the initial ideal circuit depth (after decomposition). Whenever there are overhead comparisons, we use a relative performance metric. Specifically, the relative gate ovehead (\textit{rGO}) and relative depth overhead (\textit{rDO}) is calculated as: (\textit{results\textsubscript{a}} $-$ \textit{results\textsubscript{b}}) / \textit{results\textsubscript{a}}.

Furthermore, the physical initialization of qubits follows a checkerboard pattern and with one-to-one virtual-to-physical qubit allocation, similar to the ones used in \cite{SpinQ}. Although the checkerboard pattern has a qubit density of 50 percent, in Section \ref{Densities} particularly, we use a different physical placement to achieve a range of qubit densities.

Finally, both algorithms are benchmarked with randomly generated circuits as well as eleven well-known quantum algorithms. Regarding the randomly generated quantum circuits, they consist of two-qubit and shuttle-based Z gates in three different ratios (25, 50, and 75 two-qubit gate percent), and are sampled ten times for each data point. With this choice of random algorithms, which contain many combinations of the two gate types parallelized in different ways, we aim to stress-test both routing algorithms. As for the number of gates, we fixed it at 3000 gates since its impact on the gate and depth overhead is negligible, according to \cite{SpinQ}. The eleven real quantum algorithms were scaled up to 1000 qubits to observe their performance in Section \ref{Real algorithms}. These algorithms were taken from the Qlib \cite{lin2014qlib}, Revlib \cite{wille2008revlib}, MQT Bench \cite{quetschlich2023mqtbench} and qbench \cite{bandic2023interaction} libraries.


\subsection{Benchmarking beSnake}

This section focuses on testing beSnake's various functionalities. Specifically, we used four time limits dictating the shortest path optimization level and compared them with taking only one shortest path in Section \ref{time limit}. In Section \ref{SWAP replacement}, we test the behavior of the SWAP replacement option and observe the scaling on three SWAP fidelities while maintaining the shuttle fidelity constant. Then, in Section \ref{Behavior on more connected topologies}, we are investigating beSnake's time performance on more connected topologies, whereas, in Section \ref{Densities}, we are fixing the topology sizes but vary the qubit density.

\subsubsection{Adjusting the shortest path optimization} \label{time limit}

\begin{figure*}[htpb]
     \centering
     \begin{subfigure}[b]{0.497\textwidth}
         \centering
         \includegraphics[width=\textwidth]{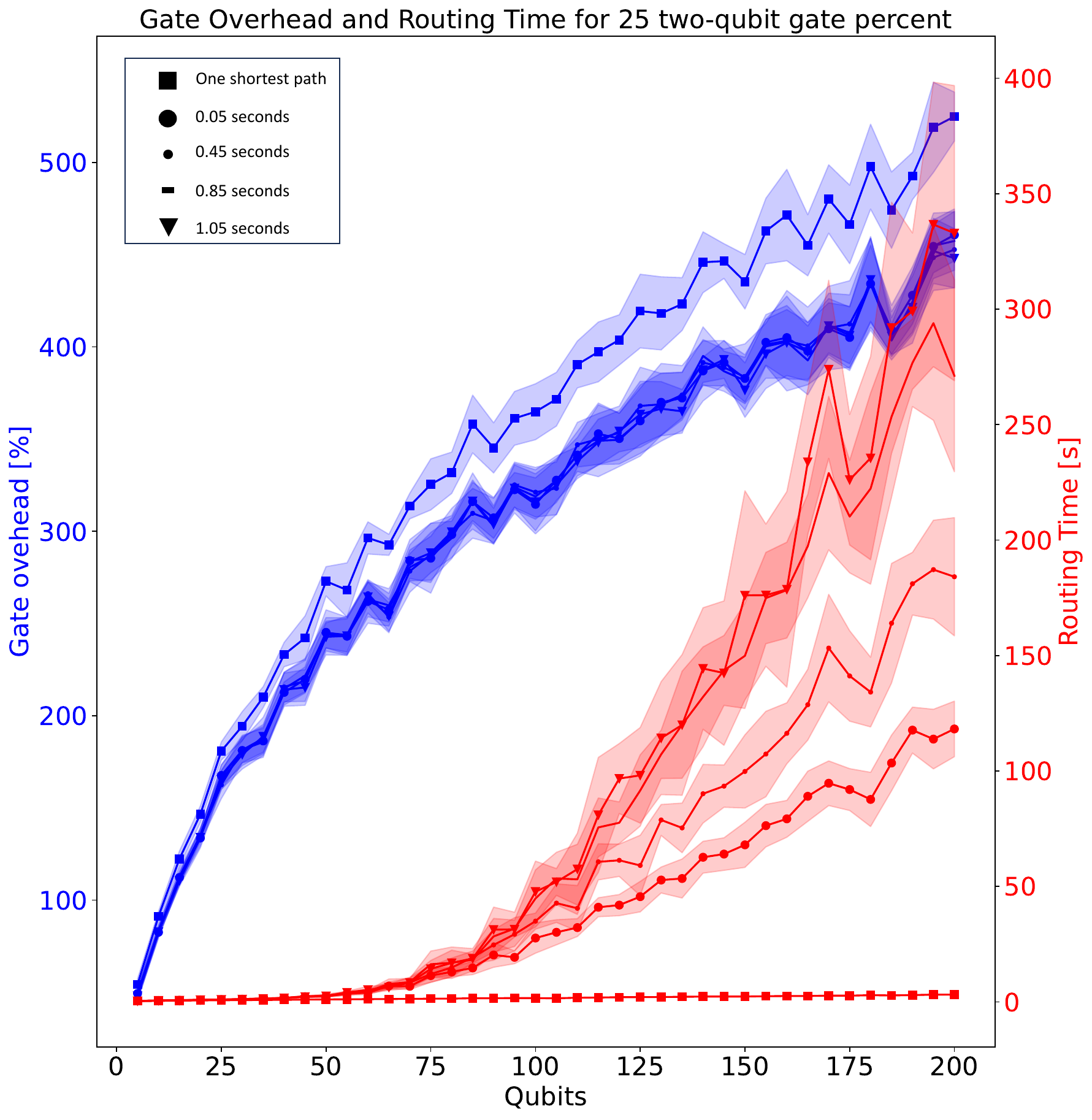}
         \caption{}
         \label{fig:time_limit_25}
     \end{subfigure}
          \begin{subfigure}[b]{0.497\textwidth}
         \centering
         \includegraphics[width=\textwidth]{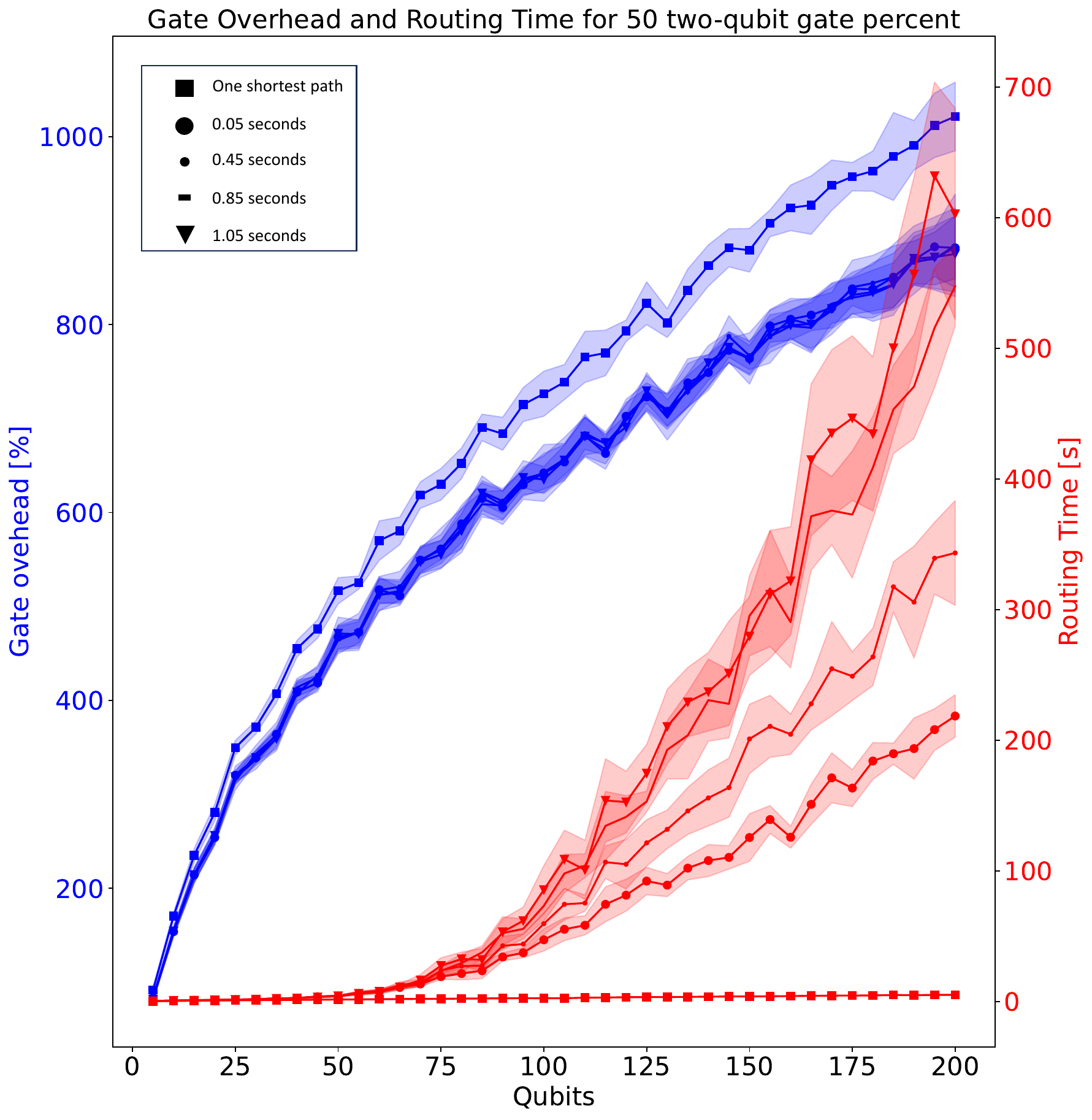}
         \caption{}
         \label{fig:time_limit_50}
     \end{subfigure}
    \begin{subfigure}[b]{0.497\textwidth}
         \centering
         \includegraphics[width=\textwidth]{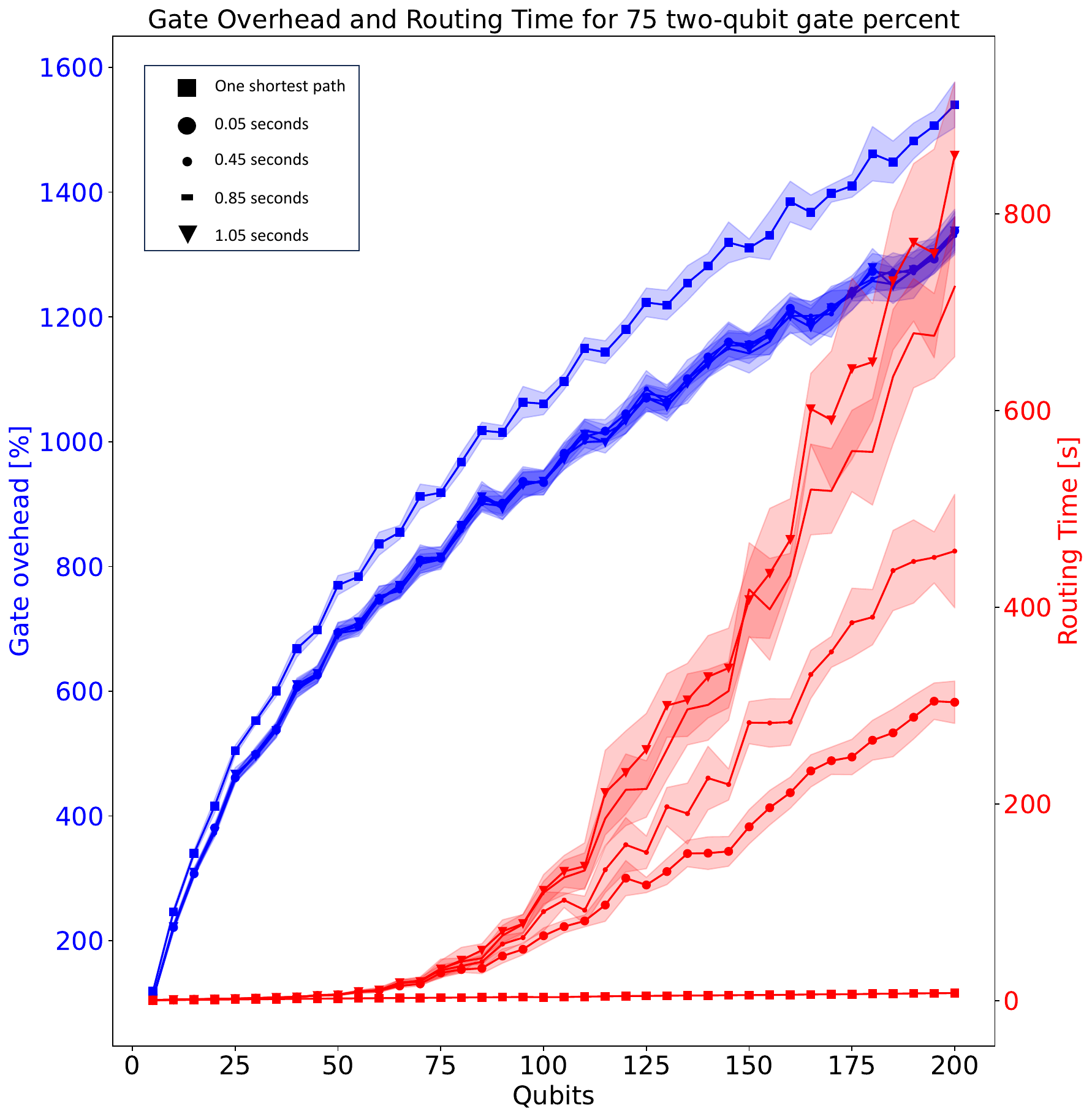}
         \caption{}
         \label{fig:time_limit_75}
     \end{subfigure}
    \begin{subfigure}[b]{0.497\textwidth}
         \centering
         \includegraphics[width=\textwidth]{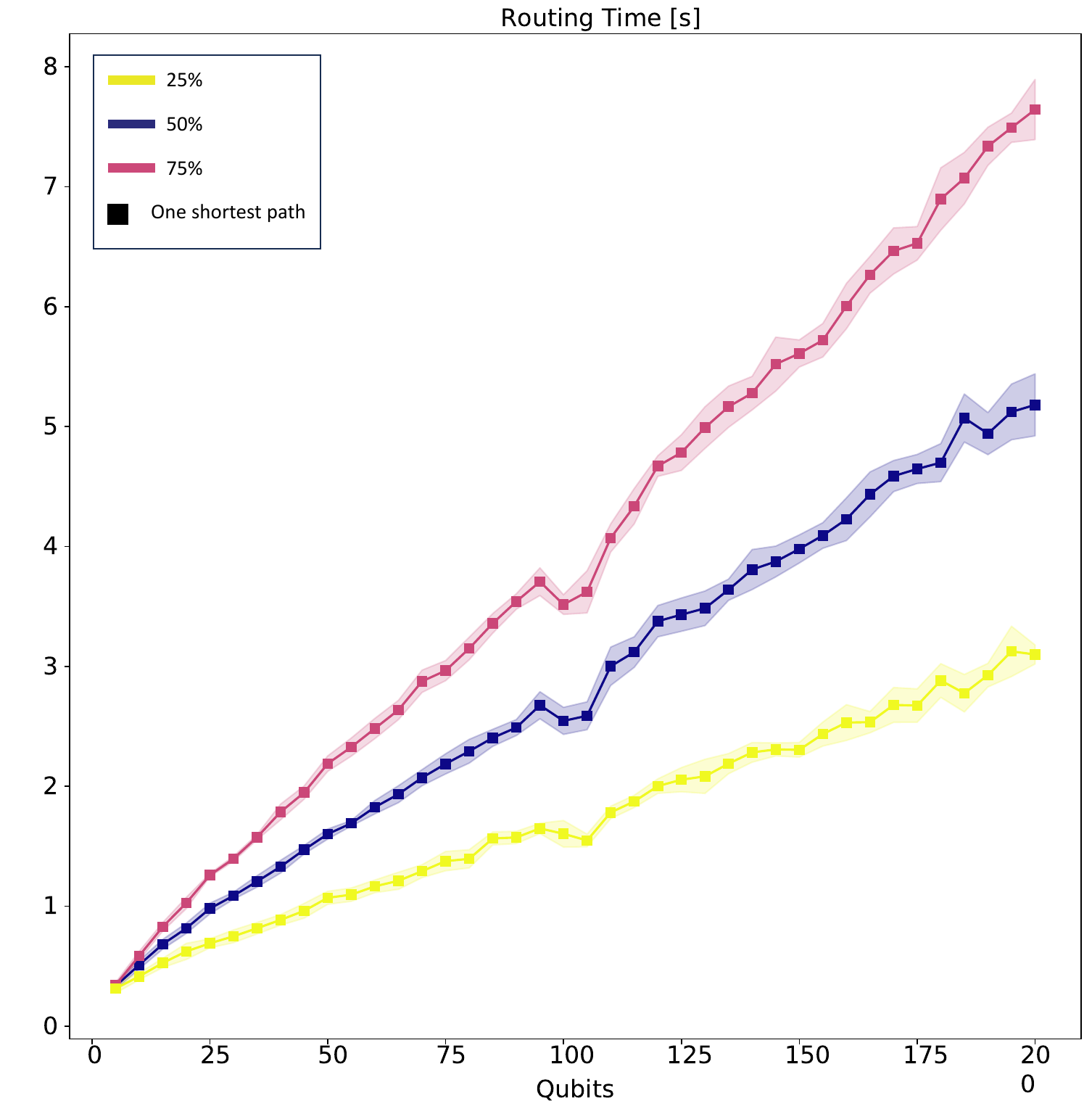}
         \caption{}
         \label{fig:time_limit_routing_time3}
     \end{subfigure}
    \caption{Routing time and gate overhead of beSnake for different shortest path optimizations. Each optimization is dictated by the time limit on finding the shortest paths, except for one option that considers just one shortest path. The first three subfigures show the results of 25, 50, and 70 two-qubit gate percentages. The last subfigure focuses only on one shortest path.}
    \label{fig:time_limit}
        
\end{figure*}

In the following simulations, we explore the effect various time limits have on gate overhead, as outlined in Section \ref{Optional functionalities}. Therefore, in Fig. \ref{fig:time_limit}, we compared the gate overhead (in blue) and total routing time (in red) for $0.05$, $0.45$, $0.85$, and $1.05$ seconds time limits imposed to \texttt{NetworkX.all\_shortest\_paths()} function against just taking one shortest path with \texttt{NetworkX.shortest\_path()}. The random circuits' gates have been fully (ideally) parallelized based only on their dependences, and beSnake is configured without the ability of SWAP replacements (see Section \ref{Optional functionalities}), such that no other factor can influence the results. The underlying question to answer with this simulation is: \newline

\begin{mdframed}[backgroundcolor=lightgreen, linecolor=black, linewidth=1pt]
\textit{\textbf{Which time limit offers the best relation between gate overhead and routing time, and how does it compare with just taking one shortest path?}} 
\end{mdframed}

In Figures \ref{fig:time_limit_25}, \ref{fig:time_limit_50} and \ref{fig:time_limit_75}, we can observe the gate overhead for 25, 50, and 75 two-qubit gate percentages, respectively. We observe that in all cases, the increased time limits yield a benefit in terms of gate overhead compared to taking only one shortest path, but the benefits are negligent among the time limits. As expected, higher two-qubit gate percentages result in higher gate overhead and routing time. Thus, the gate overhead gap between the one shortest path and the time limit simulations is more prominent in higher percentages. However, the overall routing time is severely increasing for each time limit. In fact, finding only one shortest path appears polynomial in time over the number of qubits with a small rGO increase of $11.98\%$, $11.87\%$, and $11.76\%$ on average for 25, 50, and 75 two-qubit gate percentage, respectively. To solidify this point, in Fig. \ref{fig:time_limit_routing_time3}, we isolated the one shortest path option, clearly showing a nearly linear behavior. A 0.05-second time limit is thus sufficient; increasing it beyond this limit does not yield significant gate overhead benefits, and further increases negatively impact the total routing time.

\subsubsection{SWAP replacement} \label{SWAP replacement}

\begin{figure*}[htpb]
     \centering
     \begin{subfigure}[b]{0.497\textwidth}
         \centering
         \includegraphics[width=\textwidth]{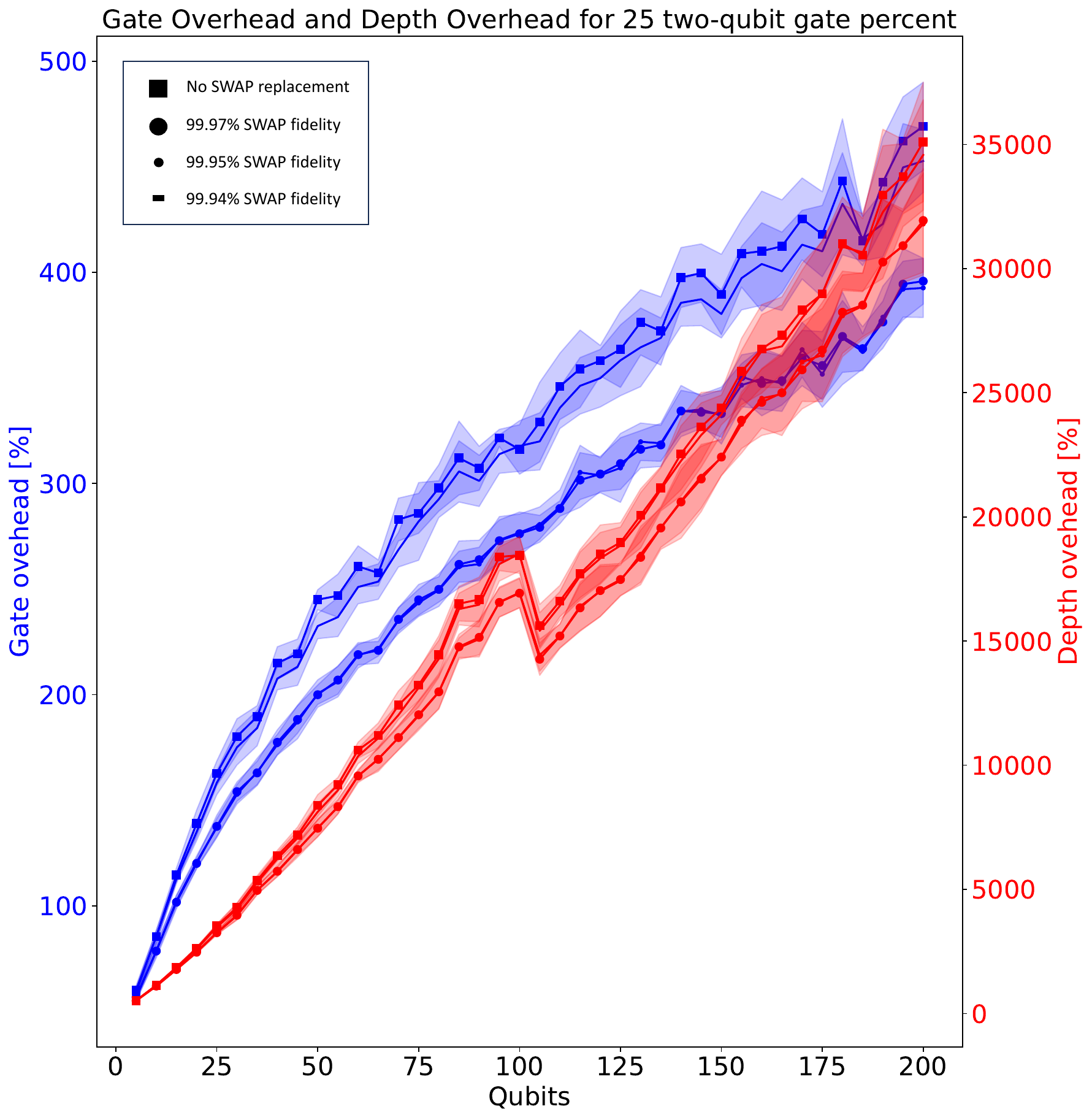}
         \caption{}
         \label{fig:swap_25}
     \end{subfigure}
     \begin{subfigure}[b]{0.497\textwidth}
         \centering
         \includegraphics[width=\textwidth]{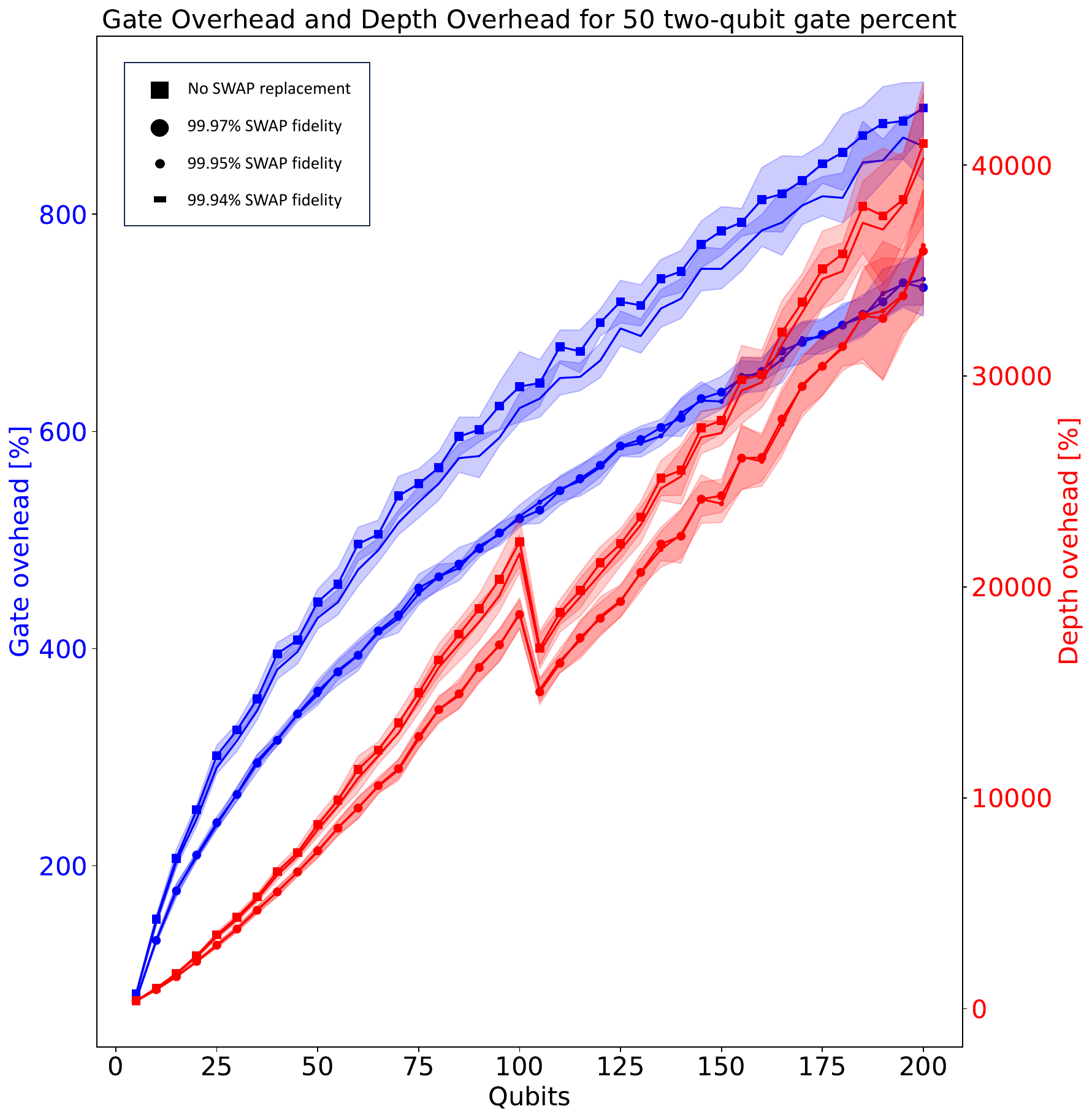}
         \caption{}
         \label{fig:swap_50}
     \end{subfigure}
    \begin{subfigure}[b]{0.497\textwidth}
         \centering
         \includegraphics[width=\textwidth]{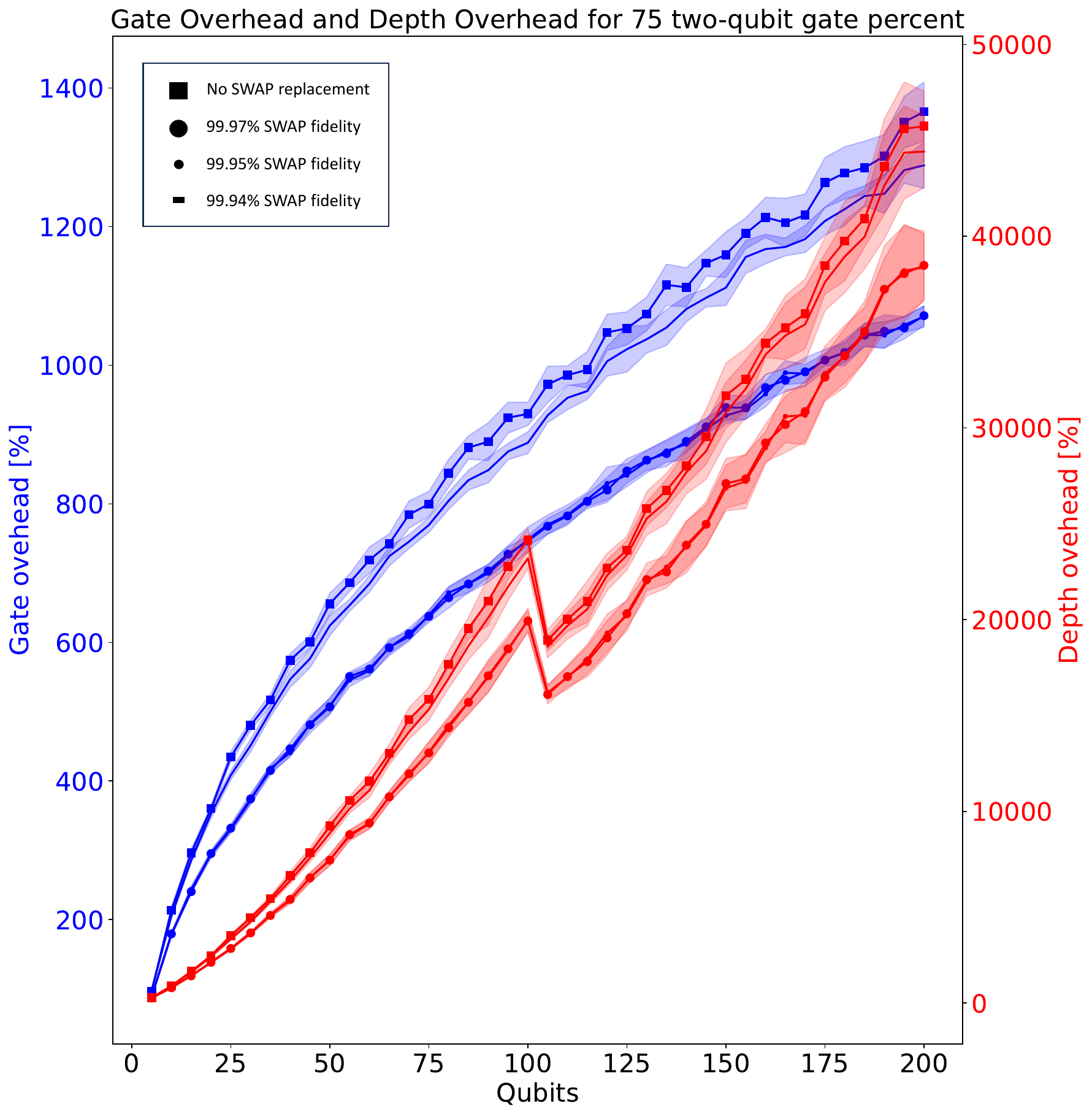}
         \caption{}
         \label{fig:swap_75}
     \end{subfigure}
          \begin{subfigure}[b]{0.497\textwidth}
         \centering
         \includegraphics[width=\textwidth]{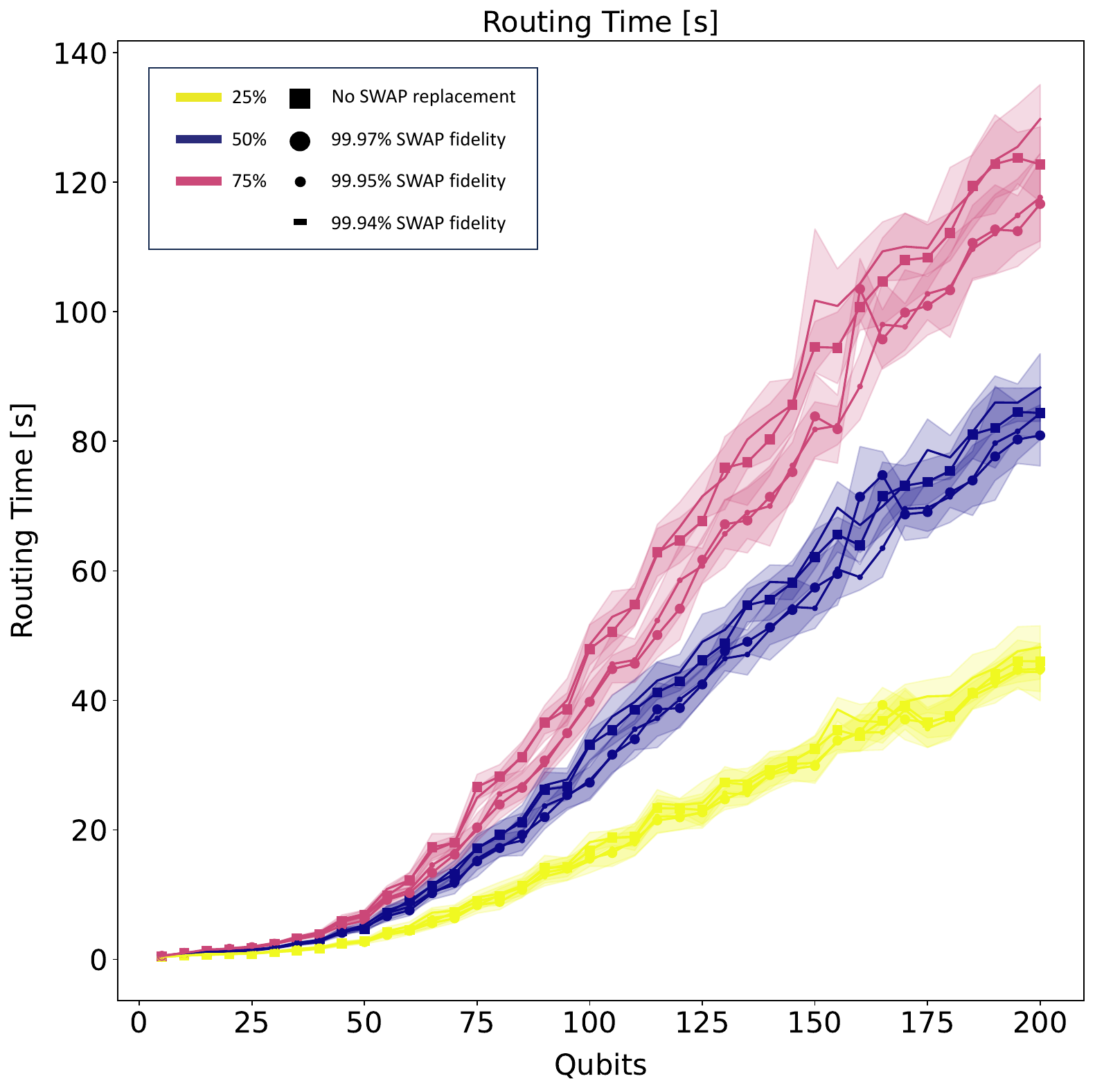}
         \caption{}
         \label{fig:swap_routing_time}
     \end{subfigure}
    \caption{Gate overhead, depth overhead, and routing time of beSnake while using the SWAP replacement option with varying SWAP fidelities. The first three subfigures show the results of 25, 50, and 70 two-qubit gate percentages, whereas the last subfigure focuses only on routing time.}
    \label{fig:swap_replacement} 
\end{figure*}

In this simulation, the behavior of the SWAP replacement option on all three metrics is investigated. To test the effect of different fidelity scales, we have fixed the shuttle to $99.98\%$, but tested SWAP gates with $99.97\%$, $99.95\%$, and $99.94\%$ fidelities. As explained in Section \ref{Optional functionalities}, the SWAP replacement is triggered when the accumulated shuttle fidelity is lower than the SWAP fidelity. Intuitively, we expect replacements to be triggered more often with higher SWAP fidelities. Finally, the gates in the random circuits are serialized in order to pronounce the effect in both overheads (when replacing multiple shuttles at once), and the shortest path optimization is fixed with a 0.05-second time limit. The underlying question to answer with this simulation is: \newline

\begin{mdframed}[backgroundcolor=lightgreen, linecolor=black, linewidth=1pt]
\textit{\textbf{Is the potential increase in routing time with beSnake worth it for architectures that support SWAP gates?}} 
\end{mdframed}

In Fig. \ref{fig:swap_replacement}, SWAP replacements improved all overheads for every two-qubit gate percentage. As expected, the SWAP replacement was utilized more often with higher SWAP fidelities and higher two-qubit gate percentages. However, there is no significant overhead difference shown in the figures for higher than $99.95\%$ (see the clustered lines), potentially due to the maximum accumulated shuttle fidelity being somewhere between $99.94\%$ and $99.95\%$. In particular, three shuttles are needed to obtain an accumulated fidelity between $99.94\%$ and $99.95\%$. This indicates that the BFS exploration never reached more than three layers, which is equivalent to three consecutive shuttles (similar to the example in Fig. \ref{fig:thirdexampleb}). Unexpectedly, we observe that the routing time has not worsened but mostly improved for higher than $99.94\%$ fidelity, possibly due to less time taken to update the intermediate representation data structure of the circuit with just one gate (a SWAP) instead of multiple shuttles. Therefore, the SWAP replacement functionality can be used without an execution time penalty in most cases. It should be noted that, given a higher qubit density on the topology, there can be more consecutive shuttles at one time, which will lower the trigger requirement and allow for even lower SWAP fidelities. 

\subsubsection{Behavior on more connected topologies} \label{Behavior on more connected topologies}

\begin{figure*}[htpb]
     \centering
     \begin{subfigure}[b]{0.497\textwidth}
         \centering
         \includegraphics[width=\textwidth]{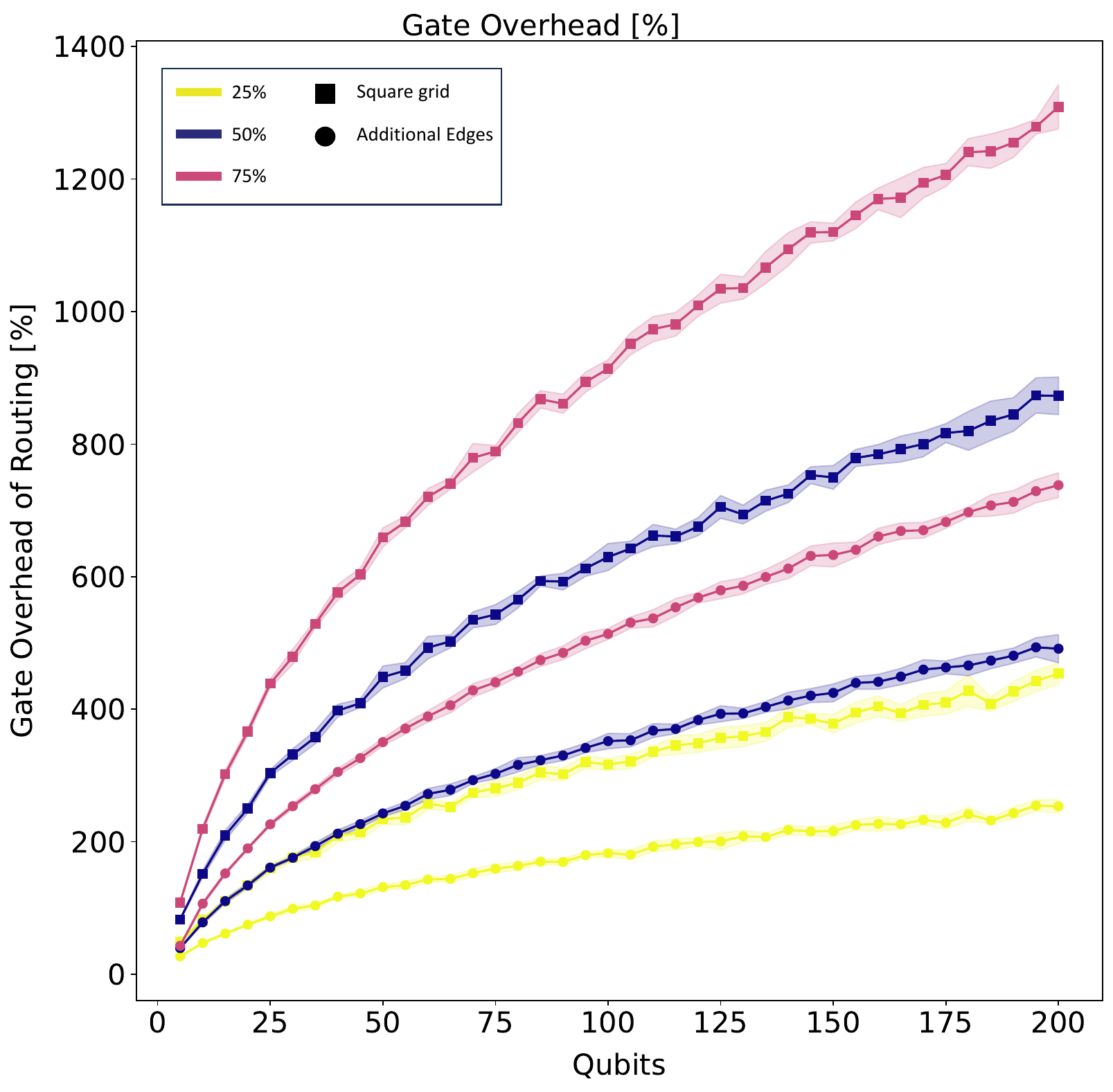}
         \caption{}
         \label{fig:more_edges_GO}
     \end{subfigure}
     \begin{subfigure}[b]{0.497\textwidth}
         \centering
         \includegraphics[width=\textwidth]{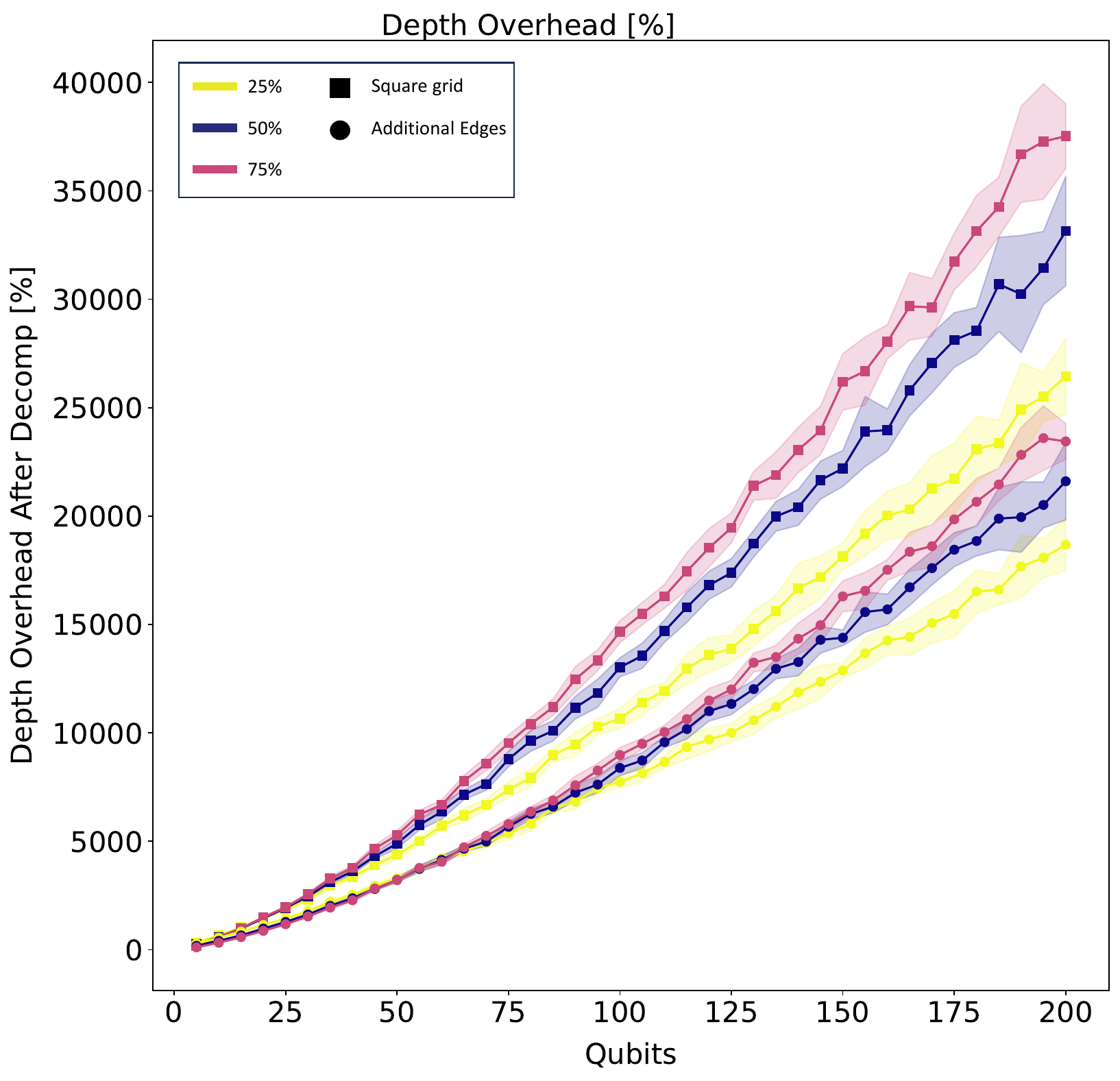}
         \caption{}
         \label{fig:more_edges_DO}
     \end{subfigure}
     \begin{subfigure}[b]{0.497\textwidth}
         \centering
         \includegraphics[width=\textwidth]{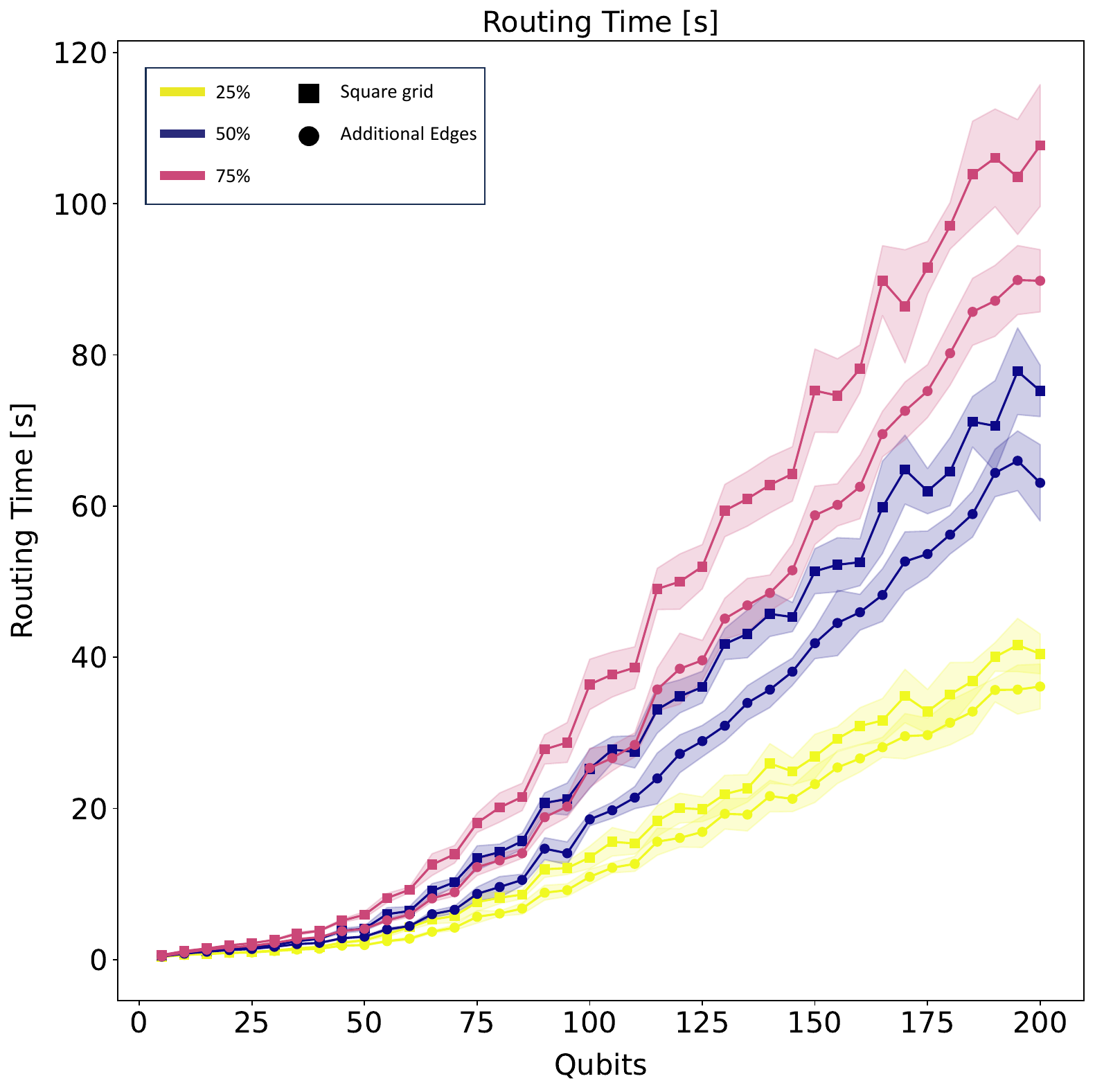}
         \caption{}
         \label{fig:more_Edges_routing_time}
     \end{subfigure}
    \caption{Gate overhead, depth overhead, and routing time of beSnake obtained on two different topologies, one square grid and one with additional diagonal edges. Each subfigure is dedicated to each of the three performance metrics.}
    \label{fig:more_edges}
\end{figure*}

In the following experiments, we are interested in seeing how beSnake performs on topologies with higher connectivity. Here, the comparison is made between a squared grid topology (i.e., each qubit is coupled with four neighboring qubits and with two at the corners) and a squared grid topology in which diagonal edges in every direction have been added (i.e., each qubit is coupled with eight neighboring qubits, and with three at the corners). Of course, it is expected that there will be an improvement in both overheads on the topologies with more connections. However, the increased graph size and solution space might negatively impact beSnake's speed in exploring solutions. The random circuits are given the same way as in Section \ref{time limit} as well as the configuration of beSnake but only with one time limit of 0.05 seconds. The underlying question to answer with this simulation is: \newline

\begin{mdframed}[backgroundcolor=lightgreen, linecolor=black, linewidth=1pt]
\textit{\textbf{Does a highly connected topology adversely affect the efficiency of beSnake in quickly finding solutions?}} 
\end{mdframed}

As can clearly be seen in Fig. \ref{fig:more_edges}, beSnake can obtain significantly better gate and depth overhead as was expected, and surprisingly, it can also achieve it in less time. In particular, beSnake is, on average, $1.17$, $1.24$, and $1.27$ faster (rTime) for 25, 50, and 75 two-qubit gate percentages, respectively. The benefit is more pronounced with higher two-qubit gate percentages, which is expected. One explanation for the faster times can be attributed to the shorter traversed paths, hence fewer steps to take and fewer BFS exploration layers.


\subsubsection{Qubit density} \label{Densities}

\begin{figure*}[htpb]
     \centering
     \begin{subfigure}[b]{0.497\textwidth}
         \centering
         \includegraphics[width=\textwidth]{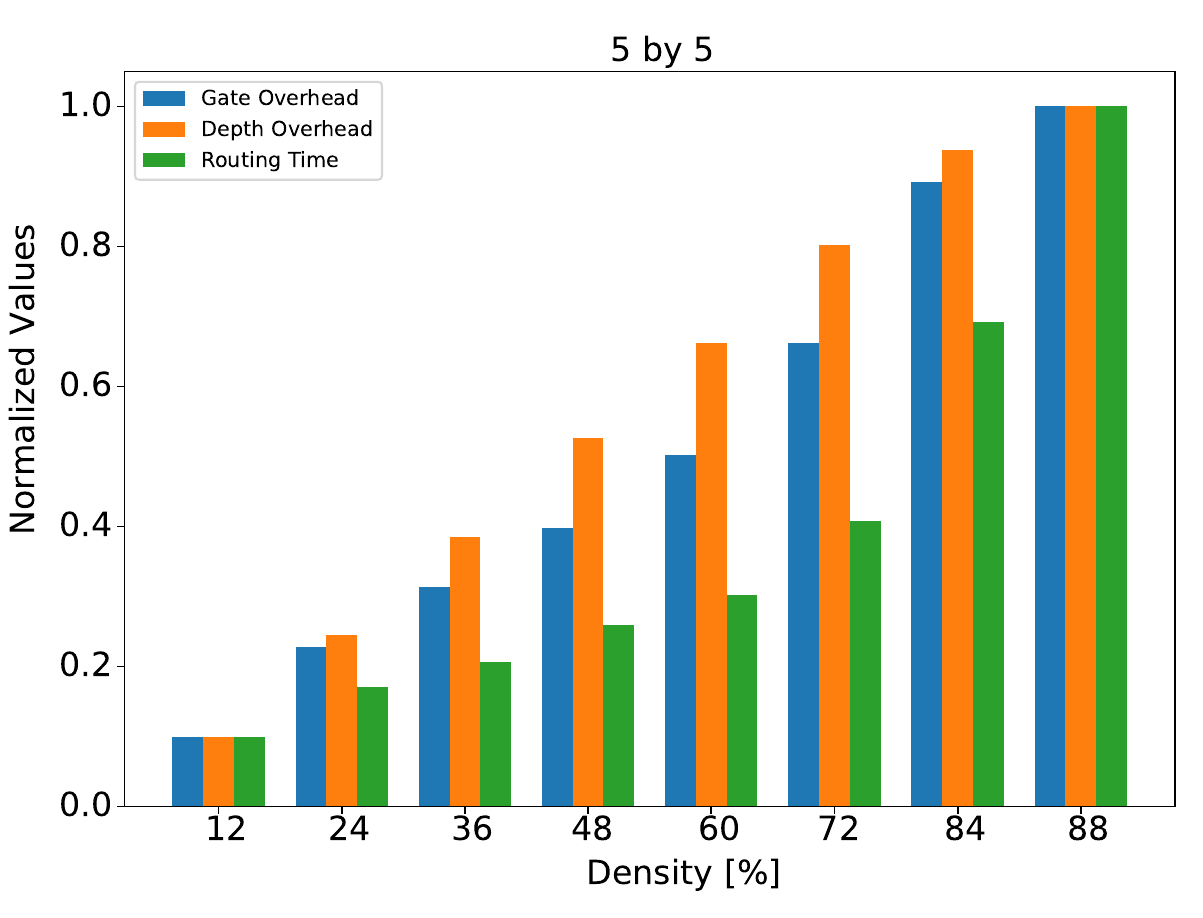}
         \caption{}
         \label{fig:5by5}
     \end{subfigure}
     \begin{subfigure}[b]{0.497\textwidth}
         \centering
         \includegraphics[width=\textwidth]{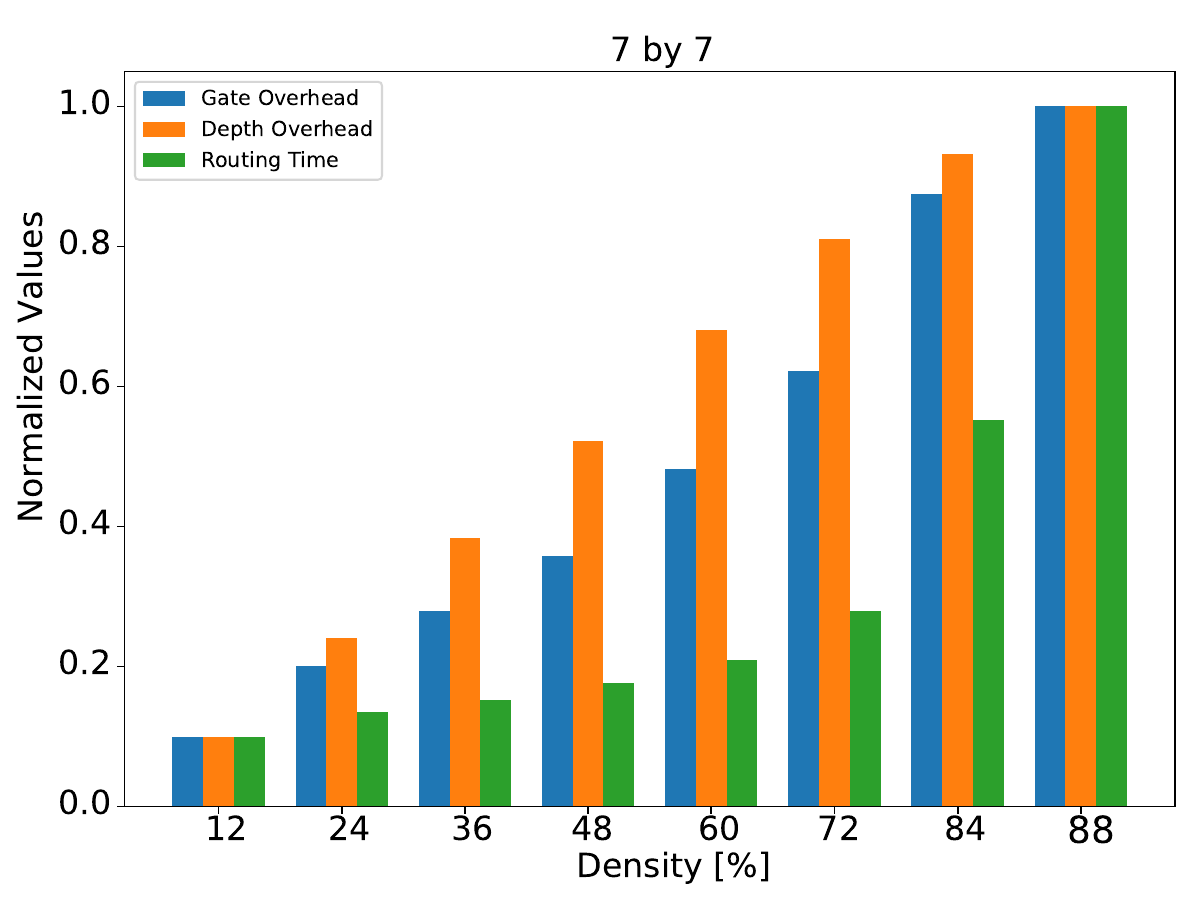}
         \caption{}
         \label{fig:7by7}
     \end{subfigure}
          \begin{subfigure}[b]{0.497\textwidth}
         \centering
         \includegraphics[width=\textwidth]{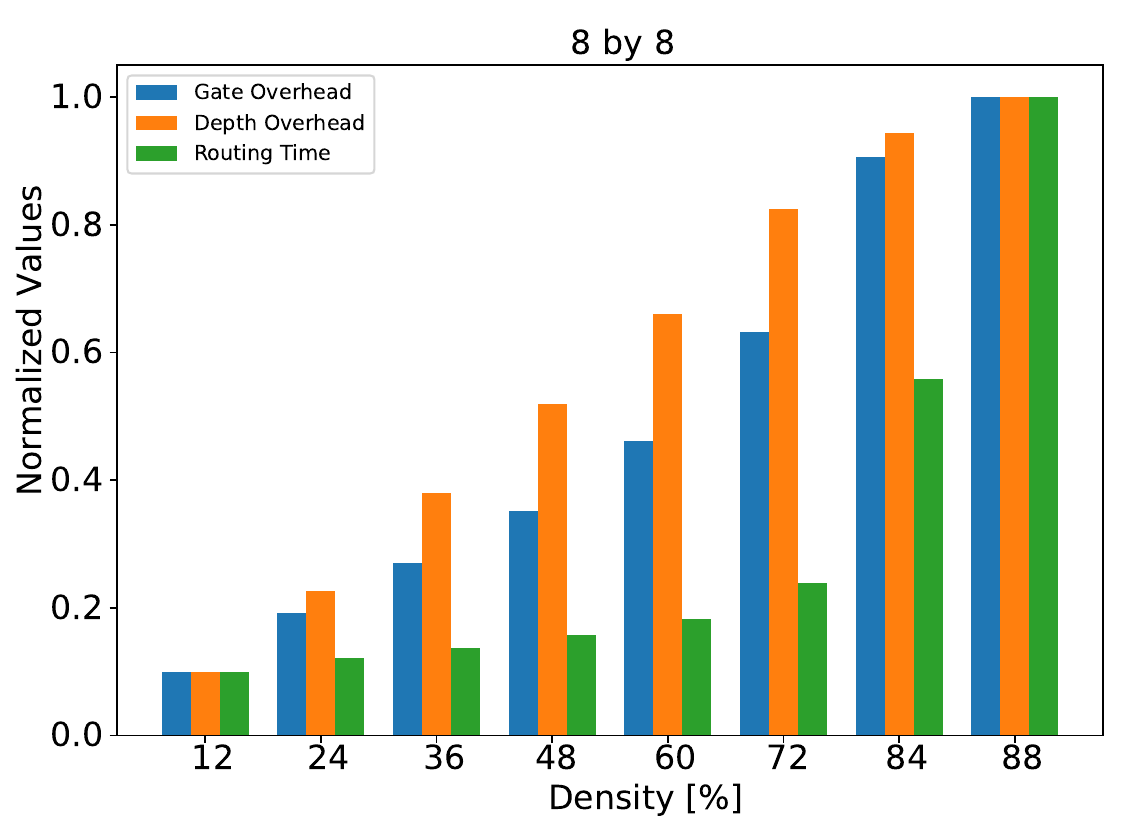}
         \caption{}
         \label{fig:8by8}
     \end{subfigure}
     \begin{subfigure}[b]{0.497\textwidth}
         \centering
         \includegraphics[width=\textwidth]{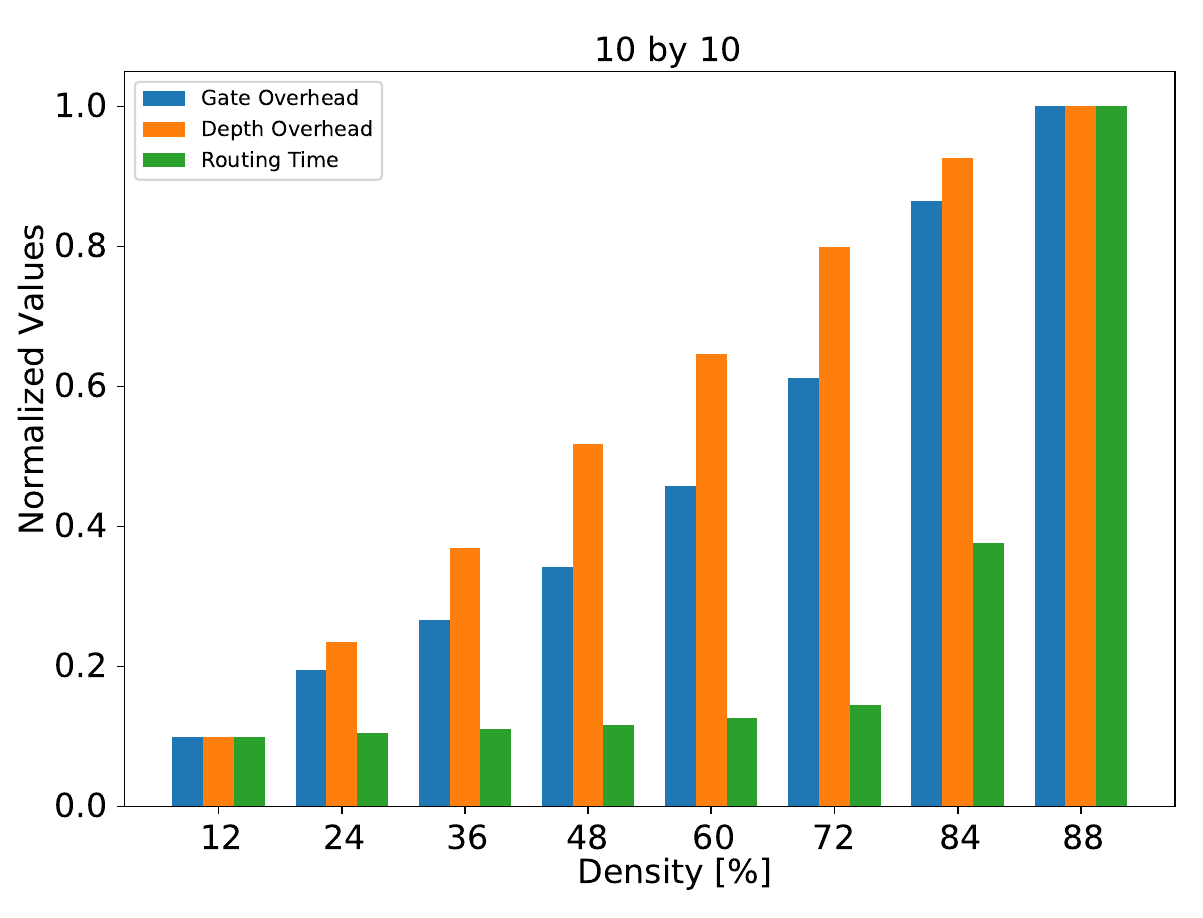}
         \caption{}
         \label{fig:10by10}
     \end{subfigure}
    \caption{beSnake's normalized performance on four fixed topology sizes in each subfigure while varying the qubit density.}
    \label{fig:densities} 
\end{figure*}  

Unlike the other simulations using a checkerboard physical initial placement pattern, we will try a different placement to vary the qubit densities here. The topology sizes are fixed, and qubits are placed next to each other in a left-to-right bottom-up approach until a particular density is reached. As for the executed circuits and beSnake's configuration, they are similar to Section \ref{time limit} with a 0.05-second time limit. The underlying question to answer with this simulation is: \newline

\begin{mdframed}[backgroundcolor=lightgreen, linecolor=black, linewidth=1pt]
\textit{\textbf{Up to which qubit density does beSnake perform in a scalable manner based on all three metrics?}} 
\end{mdframed}

In Fig. \ref{fig:densities}, we have simulated seven levels of densities in each subfigure for four different topology sizes: a 25, 49, 64, and 100 site topology. On the y axes, we can observe the normalized mean values between 0.1 and 1 for all three metrics. Focusing on the 25-site topology, routing time (in green) increases exponentially after $72\%$ density, which is expected primarily due to the additional time needed to finish each BFS exploration. This becomes more severe at $88\%$ density, where there are only $3$ empty locations on the topology. Observing the routing time for the larger topology sizes, we see a steeper upward trend after $72\%$. However, the routing time behavior until $72\%$ appears linear for all sizes. Notably, it becomes flatter for larger sizes, indicating that beSnake can perform better on larger topologies. This shows that beSnake's qubit density sweet spot is around $72\%$. 

It should be noted, however, that for another experimental setup in regards to the hardware used to run beSnake, we might obtain a different trend and possibly an improved one if beSnake becomes multithreaded, for instance. Additionally, changing device characteristics can change beSnake's performance. For example, increasing the connectivity will favor the run time based on the simulations in Section \ref{Behavior on more connected topologies}. Having said that, the total run time might be reduced equivalently for all densities, and therefore, the same behavior at $72\%$ could still appear. Whether this sweet spot (or around that region) is universally found for any given setup is likely but left open for future work.

Focusing now on both overheads, we observe overall that beSnake favors gate overhead (in blue) over depth overhead (in orange), which is a direct outcome of the core routing strategy described in Section \ref{The beSnake routing algorithm}. Then, depth overhead follows a linear trend for all sizes. Surprisingly, gate overhead creates a slight curve gravitating towards $72\%$ qubit density. Therefore, this reinforces the previous conclusion that beSnake can operate well for more than $50\%$ qubit density and particularly $72\%$ for our particular setup. Finally, the simulations show that in terms of gate and depth circuit overhead, there are no exponential penalties for higher than $72\%$ densities.

\subsection{Comparison between beSnake and SBS}

In this section, we will compare the performance of both routers for random and real circuits for up to 1000 qubits. To facilitate a fair comparison (later named as \textit{beSnake fair}), the circuits are given with their gates as parallelized as possible based on SBS's maximum capability \cite{SpinQ}. The SWAP replacement of beSnake has been deactivated as well because SBS does not support it. Additionally, beSnake's shortest-path optimizations are disabled, and we only use one shortest path (see Section \ref{time limit}) to obtain the quickest routing times at the expense of overhead (see Section \ref{time limit}). As noted before, beSnake can handle more complex routing tasks with ideally scheduled circuits based purely on their gate's dependencies, and, for comparison purposes, we will provide such results in Section \ref{Real algorithms}, named as \textit{beSnake full}.

\subsubsection{Randomly generated circuits with up to 1000 qubits} \label{Random generated circuits up to 1000 qubits}

\begin{figure*}[htpb]
     \centering
     \begin{subfigure}[b]{0.497\textwidth}
         \centering
         \includegraphics[width=\textwidth]{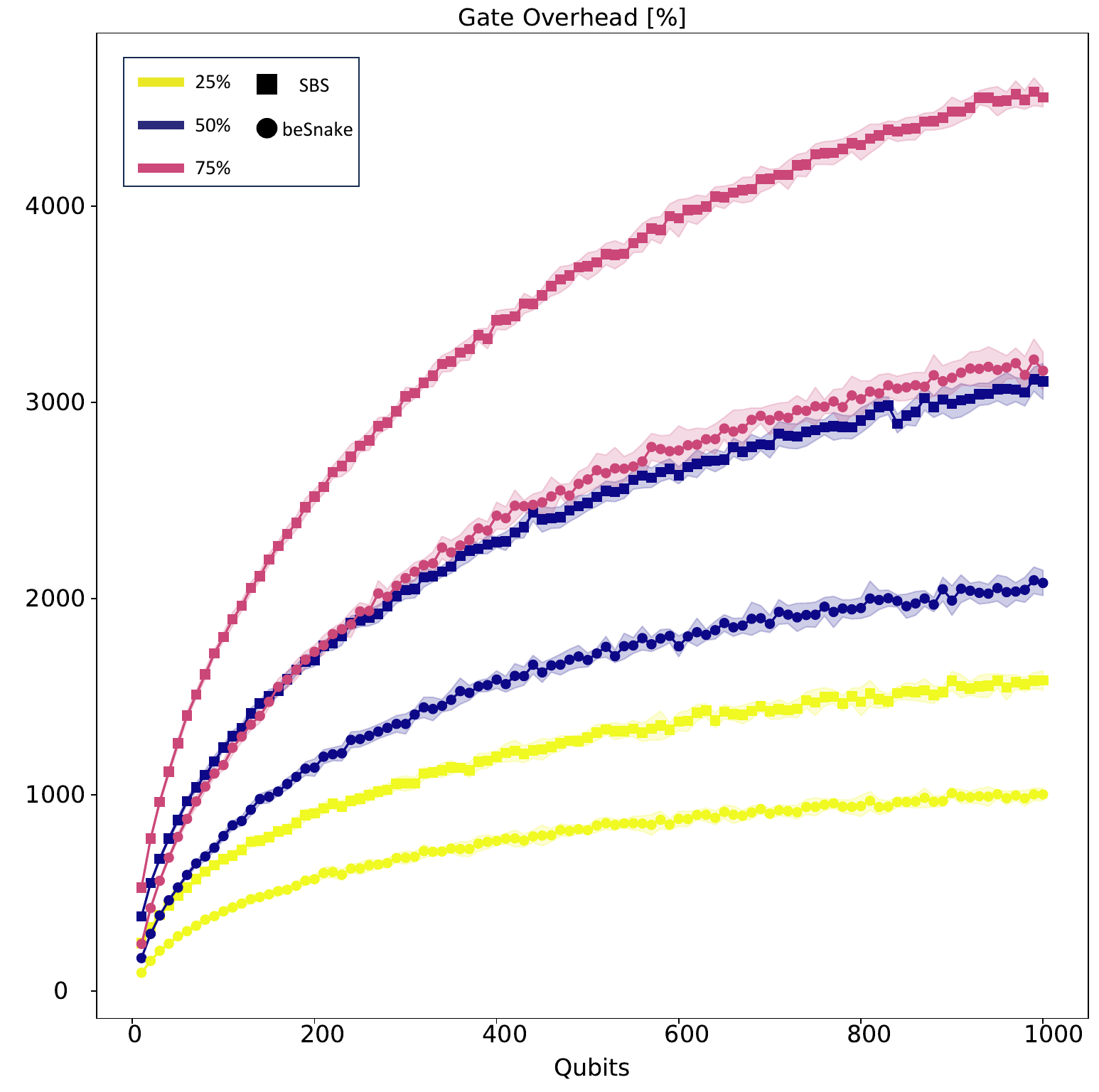}
         \caption{}
         \label{fig:GO_random_1000}
     \end{subfigure}
     \begin{subfigure}[b]{0.497\textwidth}
         \centering
         \includegraphics[width=\textwidth]{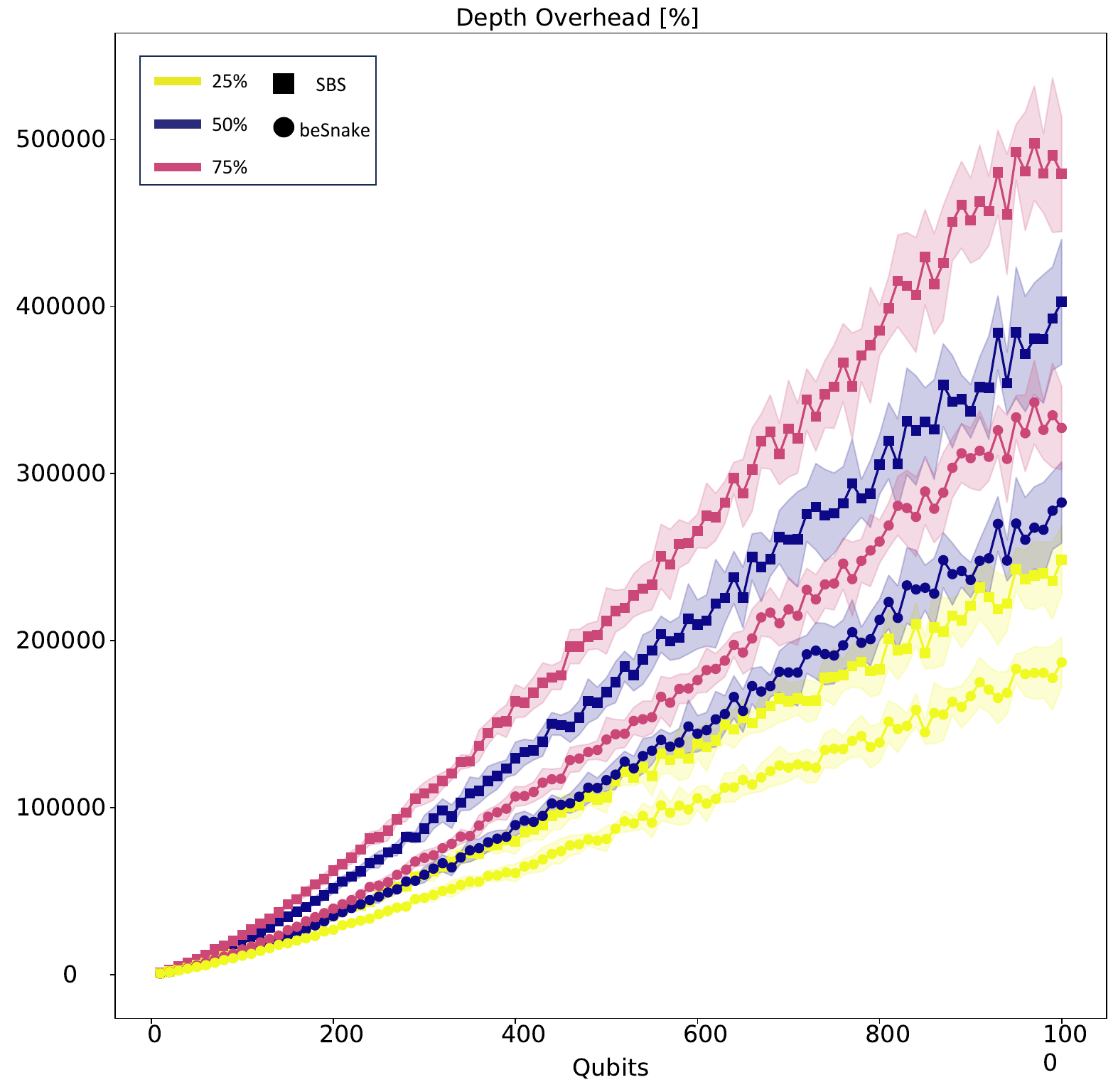}
         \caption{}
         \label{fig:DO_random_1000}
     \end{subfigure}
     \begin{subfigure}[b]{0.497\textwidth}
         \centering
         \includegraphics[width=\textwidth]{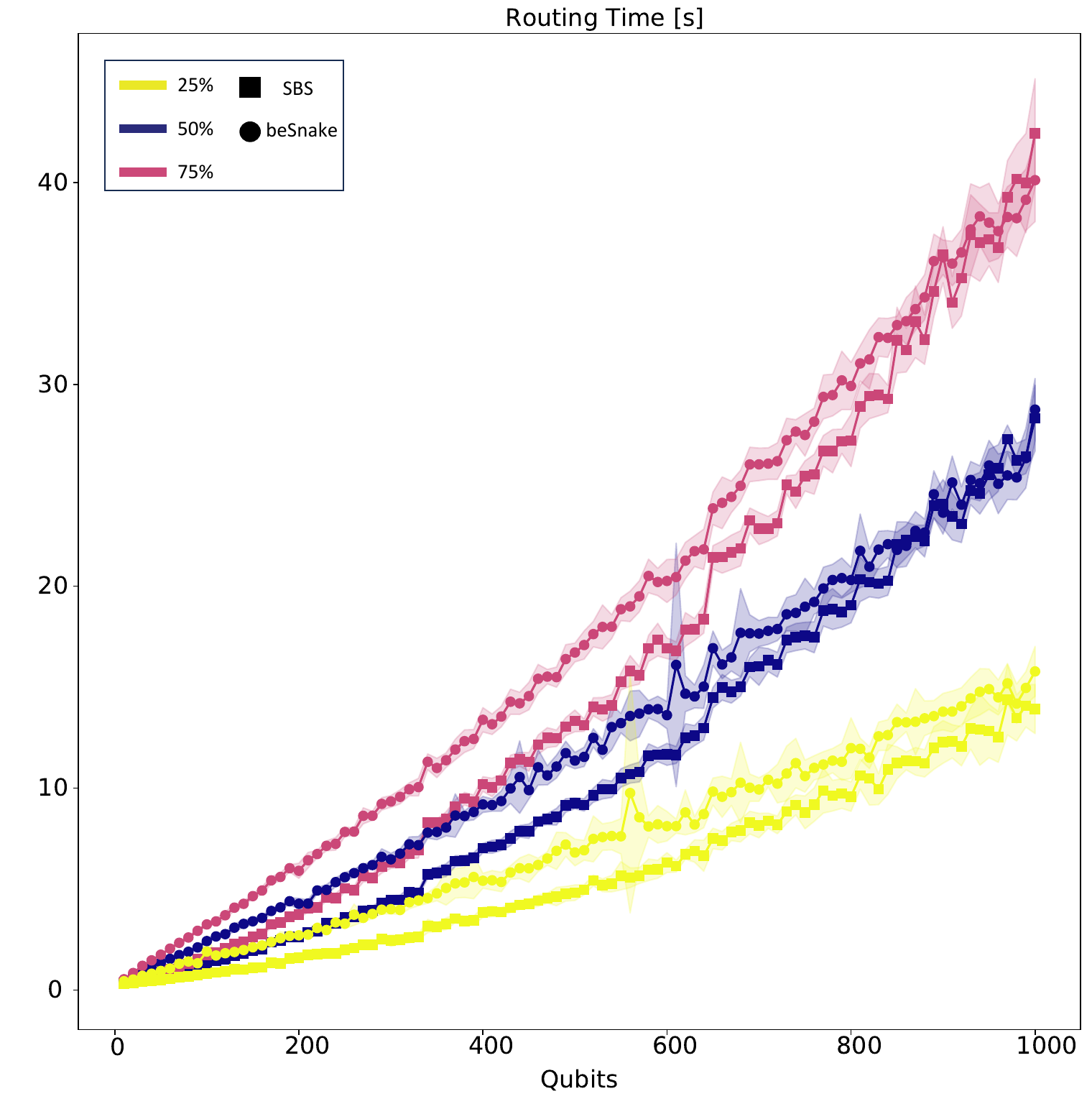}
         \caption{}
         \label{fig:Routing_time_1000}
     \end{subfigure}
    \caption{Performance comparison between beSnake (circle markers) and SBS (square markers) on random algorithms with up to 1000 qubits.}
    \label{fig:random_1000_qubits}
\end{figure*}

The underlying question to answer with this simulation is: \newline

\begin{mdframed}[backgroundcolor=lightgreen, linecolor=black, linewidth=1pt]
\textit{\textbf{Which routing algorithm will more likely produce lower gate and circuit depth overhead for scalable architectures that go up to 1000 qubits, and at what relative time cost?}} 
\end{mdframed}

In Figures \ref{fig:GO_random_1000} and \ref{fig:DO_random_1000}, the gate overhead and depth overhead of both algorithms (marked as \textit{SBS} and \textit{beSnake}) are shown for the different two-qubit gate percentages (in yellow, blue, pink). We can observe beSnake (circle markers) a relative improvement in gate overhead (rGO) of $32.2\%$ and in depth overhead (rDO) of $30.3\%$ on average compared to SBS (square markers). Additionally, we observe beSnake's relative benefit in both overheads increasing with higher qubit counts. Looking at the three two-qubit gate percentages, we see an increased gate overhead improvement with higher percentages (see the distance between curves of the same color). More specifically, the average rGO is at $36.61$\% for 25\%, $32.64\%$ for 50\%, and $30.36\%$ for 75\%, and the average rDO is calculated at $24.05\%$ for 25\%, $30.5\%$ for 50\%, and $33.19$\% for 75\%. beSnake will only be, on average, $0.78$ slower (rTime) than SBS at 25\%, $0.88$ at 50\%, and $0.87$ at 75\%. Notably, both algorithms' routing time starts getting closer around and after 900 qubits, indicating that beSnake takes almost the same amount of time at 1000 qubits, especially for more complex routing tasks (i.e., 75 two-qubit gate percentage shown in Fig. \ref{fig:Routing_time_1000}). Considering the performance gains, beSnake is a more attractive solution for large-scale architectures overall. Finally, as discussed before, in these comparison simulations, beSnake is not utilized fully; hence, we expect even better performance.

\subsubsection{Real algorithms with up to 1000 qubits} \label{Real algorithms}

\begin{table*}[htpb]
\caption{Relative performance of beSnake's two configurations, \textit{beSnake fair} and \textit{beSnake full}, over SBS for real algorithms scaled up to 1000 qubits.}
\includegraphics[width=1\textwidth]{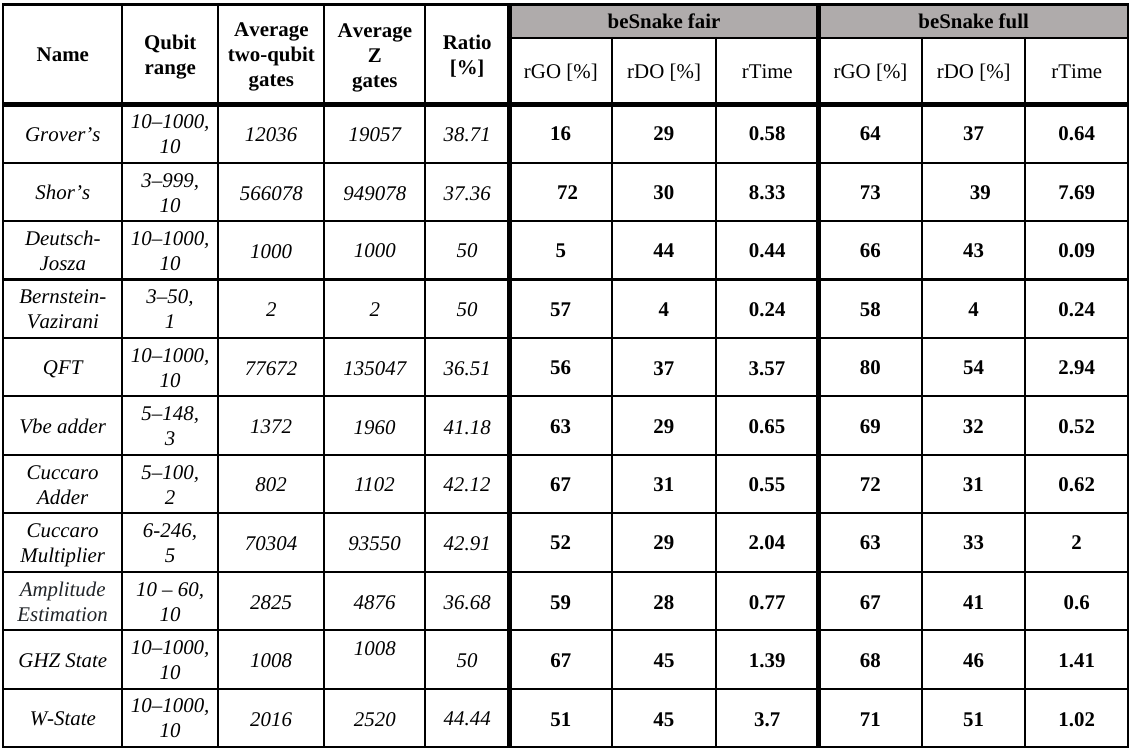}
    \parbox{\textwidth}{\textbf{Ratio}: (Average of two-qubit gates + Average of Z gates) / Average of two-qubit gates. \textbf{beSnake fair}: beSnake is configured such that it matches the maximum capabilities of the shuttle-based SWAP (SBS) algorithm. \textbf{beSnake full}: beSnake is used at its full capacity. \textbf{rGO [\%] and rDO [\%] } refer to relative gate and depth overhead, respectively: (\textit{SBS average results} $-$ \textit{beSnake average results}) / \textit{SBS average results}. \textbf{rTime}: \textit{SBS average time} / \textit{beSnake average time}. }
    \label{fig:real_table}
\end{table*}

In the following simulations, we compare the performance of SBS over beSnake for real algorithms with up to 1000 qubits. Previously, the random circuit set was produced to stress test the capabilities of both routing algorithms; however, the results are not representative of routing for real algorithms, which might exhibit different circuit characteristics. Similarly to Section \ref{Random generated circuits up to 1000 qubits}, we will compare both routers fairly to see their relative performance. However, we will also assume ideally parallelized circuits and enable the SWAP replacement option (at 99.95\% SWAP fidelity) in order to test beSnake at its full capacity. The underlying question to answer with this simulation is: \newline

\begin{mdframed}[backgroundcolor=lightgreen, linecolor=black, linewidth=1pt]
\textit{\textbf{How much better can beSnake perform over SBS on real quantum algorithms, and is it worth the potentially increased routing time?}} 
\end{mdframed}

Table \ref{fig:real_table} summarizes the relative performance of beSnake over SBS (used as a baseline) in two configurations, one fair comparison, \textit{beSnake fair}, and one scenario with beSnake at its full capacity, \textit{beSnake full}. Due to algorithm availability, we have specified each algorithm's qubit range and incremented qubit steps on the second column. One the next 3 columns, we have provided the average two-qubit and Z gates together with their ratio. However, it should be noted that the number or percentage of gates is not the only performance indicator or predictor. To get a full picture of their features, a deeper analysis has to be conducted with the help of the algorithm's internal interaction and dependency graph \cite{SpinQ,bandic2023interaction}. In this simulation, we are focusing on the performance comparison of the two routers rather than on specific quantum algorithm performance or how these algorithms scale. 

Moving on to the rest of the columns, we provide the relative performance of beSnake over SBS for gate overhead, depth overhead, and routing time as defined in Section \ref{Experimental Setup}. For example, beSnake obtained a 16\% gate overhead and 29\% depth overhead improvement over SBS under a fair comparison for Grover's algorithm. As for routing time, beSnake is 42\% slower (\textit{rTime} = 0.58) than SBS. 

Looking only at the fair comparisons overall, we observe an improvement in both rGO and rDO across all algorithms, with up to a 67\% improvement in rGO and up to 45\% improvement in rDO. On the one hand, rTime results for Deutsch-Josza (up to 1000 qubits) and for Bernstein-Vazirani (up to 50 qubits) seem relatively high for their low average two-qubit and Z gate values. Evidently, these algorithms do not require a lot of routing; therefore, the relative time increase possibly comes from beSnake's code initialization overhead (i.e., loading libraries, allocating memory, initializing variables, and setting up data structures) and not the actual routing process. On the other hand, Cuccaro Multiplier and QFT, two highly connected circuits \cite{SpinQ}, obtained more than 50\% rGO and rDO improvement with more than 2 and 3,5 times faster routing time. 

Focusing on \textit{beSnake full} columns, we see the biggest improvement in rGO, and not in rDO, compared to \textit{beSnake fair}, and in a few algorithms, this was achieved with less routing time. This indicates that beSnake can handle highly parallelized circuits faster than less parallelized ones despite imposing more demanding routing tasks, and it can always do so with less gate overhead. This shows that, on average, holding in place qubits to satisfy two-qubit gates, as seen in the third example in Section \ref{Third Example: Satisfying parallelized gates}, does not negatively impact the gate overhead. This can be attributed to increased available space in larger topologies, which offers more alternative shortest paths around the fixed positions. Another contributing factor to this gate overhead benefit is the SWAP replacement option. Regardless, \textit{beSnake full} can reduce both overheads more compared to \textit{beSnake fair} for up to 1000 qubits while taking less time in some real algorithms. 

\section{Conclusion} \label{Conclusions}

This paper presents beSnake, a novel routing algorithm designed explicitly for scalable spin-qubit architectures. beSnake addresses the unique challenges posed by spin-qubit systems, such as the incorporation of the shuttle operation over the traditional SWAP gate used in other qubit technologies. We presented beSnake's ability to dynamically adapt to different routing challenges and efficiently handle complex scenarios involving multiple parallel gates. We stress-tested beSnake in various configuration options to provide valuable performance insights. In particular, we showed that adjusting the heuristic of the shortest path selection can offer a significant speed improvement with a relatively small overhead cost. Then, beSnake was tested on its SWAP replacement option and its capacity to handle a highly connected topology faster than a less connected one. Furthermore, we gave an insight into the optimal qubit density, up to $72\%$, that beSnake can route in a scalable manner. In the second simulation phase, we conducted an extensive comparison between the SBS routing algorithm acting on random and real algorithms with up to 1000 qubits, in which beSnake demonstrated significant improvements, with up to $80\%$ and $54\%$ on average in gate overhead and depth overhead, respectively, and up to $8.33$ times faster routing time.

While performing our research, several avenues for future work have emerged, which warrant further exploration and investigation. Regarding the mechanisms for dealing with blockages discussed in \ref{Mechanisms for blockades}, the list of alternative approaches can be further expanded for such edge cases. This should be done with scalability in mind in terms of the extra computation time and whether it will be worth it based on their utility on the particular topological patterns and size of the coupling graph. One example is to move qubits for two or more gates concurrently, or to allow the operand qubits to be pushed by other qubits. This will possibly improve the output circuit's overhead but also make it slower for beSnake to find a solution. In other cases, it can increase both overheads because of deeper BFS explorations due to path conflicts within the different layers.

As for ways to sort parallelized Z and two-qubit gates contained in a cycle, besides the one discussed in \ref{Handling shuttle-based Z rotations}, more nuanced techniques will most likely increase the algorithm's complexity but improve both overheads. One possible direction is to create a heuristic upon which the order of all the different gate types is sorted based on their predicted routing paths. Regardless, such strategies and those discussed previously should be explored and compared in targeted architectures that will use them.

Last but not least, as mentioned, beSnake utilizes NetworkX \cite{SciPyProceedings_11}. However, similar libraries are known to be more time and memory-efficient, especially for large graphs with complex structures \cite{leskovec2016snap,staudt2016networkit}. This is attributed to NetworkX's pure Python implementation compared to other libraries written in C/C++. NetworkX was chosen for its user-friendly API interface while offering flexibility for future adaptations. As quantum devices increase in size beyond 1000 qubits, better performance might be gained by using other libraries, such as igraph \cite{csardi2006igraph}, graph-tool \cite{peixoto_graph-tool_2014}, or rustworkx \cite{treinish2021rustworkx}.

\bibliographystyle{IEEEtran}
\bibliography{bibliography}

\end{document}